%% file: main.tex
\documentclass[manuscript, screen,table]{acmart} 
\usepackage{booktabs} 
\usepackage{mdframed}
\mdfdefinestyle{theoremstyle}{%
skipabove=6pt,linecolor=black,linewidth=0.5pt,%
frametitlerule=true,backgroundcolor=gray!10,%
frametitlebackgroundcolor=gray!30,
roundcorner=4pt,innertopmargin=\topskip,
}
\mdtheorem[style=theoremstyle]{definition}{Definition}
\usepackage{mathtools}

\usepackage{setspace}
\usepackage{enumitem}
\usepackage{multirow}
\usepackage{tcolorbox,tikz}
\usepackage{algorithm, algorithmic}
\usepackage{mathtools}
\usepackage[font=small,labelfont=bf]{caption}
\usepackage{subcaption}
\usepackage{csquotes}
\usepackage[graphicx]{realboxes}

\acmJournal{TOSEM}

\begin{document}
\title{OSS effort estimation using software features similarity and developer activity-based metrics}
\author[1]{Ritu Kapur} 
\author[2]{Balwinder Sodhi}
\affiliation{%
  \institution{Indian Institute of Technology Ropar}
  \department{Department of Computer Science and Engineering}
  \streetaddress{P.O. Box 1212}
 \city{Rupnagar}
 \state{Punjab}
\postcode{140001}
\country{India.}
}

 \renewcommand\shorttitle{SDEE using software features similarity and VCS metrics}
\renewcommand\shortauthors{R. Kapur and B. Sodhi}
 
 \email[1]{dev.ritu.kapur@iitrpr.ac.in}
 \email[2]{sodhi@iitrpr.ac.in}

\begin{abstract}
Software development effort estimation (SDEE) generally involves leveraging the information about the effort spent in developing similar software in the past. Most organizations do not have access to sufficient and reliable forms of such data from past projects. As such, the existing SDEE methods suffer from low usage and accuracy.

We propose an efficient SDEE method for open source software, which provides accurate and fast effort estimates. The significant contributions of our paper are i) Novel \emph{SDEE software metrics} derived from developer activity information of various software repositories, ii) \emph{SDEE dataset} comprising the SDEE metrics' values derived from $\approx13,000$ GitHub repositories from 150 different software categories, iii) an effort estimation tool based on SDEE metrics and a \emph{software description similarity model}. Our software description similarity model is basically a machine learning model trained using the  Paragraph Vectors algorithm on the software product descriptions of GitHub repositories. Given the software description of a newly-envisioned software, our tool yields an effort estimate for developing it. 

Our method achieves the highest Standard Accuracy score of 87.26\% (with cliff's $\delta$=0.88 at 99.999\% confidence level) and 42.7\% with the Automatic Transformed Linear Baseline model. Our software artifacts are available at  \url{https://doi.org/10.5281/zenodo.5095723}.

\end{abstract}

%
%


%
%


\maketitle

\include{body}

\include{appendix}
\end{document}

%% file: body.tex
\section{Introduction}

Software Development Effort Estimation (SDEE) for a newly envisioned software project is performed based on the experiential knowledge gained from software developed in the past. The sources of such \enquote{past projects} are limited to the development team and the organization procuring the new software. Also, the existing SDEE methods are either dependent on experts or are inefficient and slow. Therefore, it is desirable to have tools and techniques which help to provide accurate software development effort estimates on time. The Open Source Software (OSS) repository hosts, such as GitHub, are rich software development  sources for various software projects.

 \begin{definition}[Software project description]
By \textbf{software project description}, we refer to the main description document available at the software's GitHub repository.  Such a file typically comprises the following details:
\begin{enumerate}
    \item \emph{Brief description of the software}: A summary of its functionality and characteristics. For instance, the software might be a library, plugin, or framework.  
    \item \emph{Functionality}: It describes \emph{what} the software does. The  description of all the primary functions performed by the software. For instance, PMD\footnote{\url{https://github.com/pmd/pmd}} -- a source code analyzer tool, detects common programming flaws such as unused variables, empty catch blocks, and unnecessary object creation.
    \item \emph{Supporting platform}: It lists the supportable operating systems  and programming platforms.    
    \item \emph{Execution-specific details}: These include the essential information about the script and steps involved in running the software. 
\end{enumerate}
\end{definition}
\begin{definition}[Paragraph Vector Algorithm]
\label{def:pva}
\textbf{Paragraph Vector Algorithm (PVA)} is an unsupervised machine learning (ML) algorithm that learns fixed-length feature representations from variable-length pieces of texts, such as sentences, paragraphs, and documents. The algorithm represents each document by a dense vector trained to predict words in the document \cite{le2014distributed}.
\end{definition}
 
Stated broadly, the objective of the work presented in this paper is
\enquote{\emph{Given the software description of a newly-envisioned software, estimate the software development effort required to develop it.}} To estimate the effort, we leverage the developer-activity information of the existing similar GitHub repositories. To detect the similarity among software, we develop a software similarity detection model by training PVA  on the software product descriptions taken from the OSS projects available at Version Control System (VCS) repositories, such as GitHub\footnote{\url{https://github.com/}}. 

\begin{figure*}
\centering
\includegraphics[width=120mm]{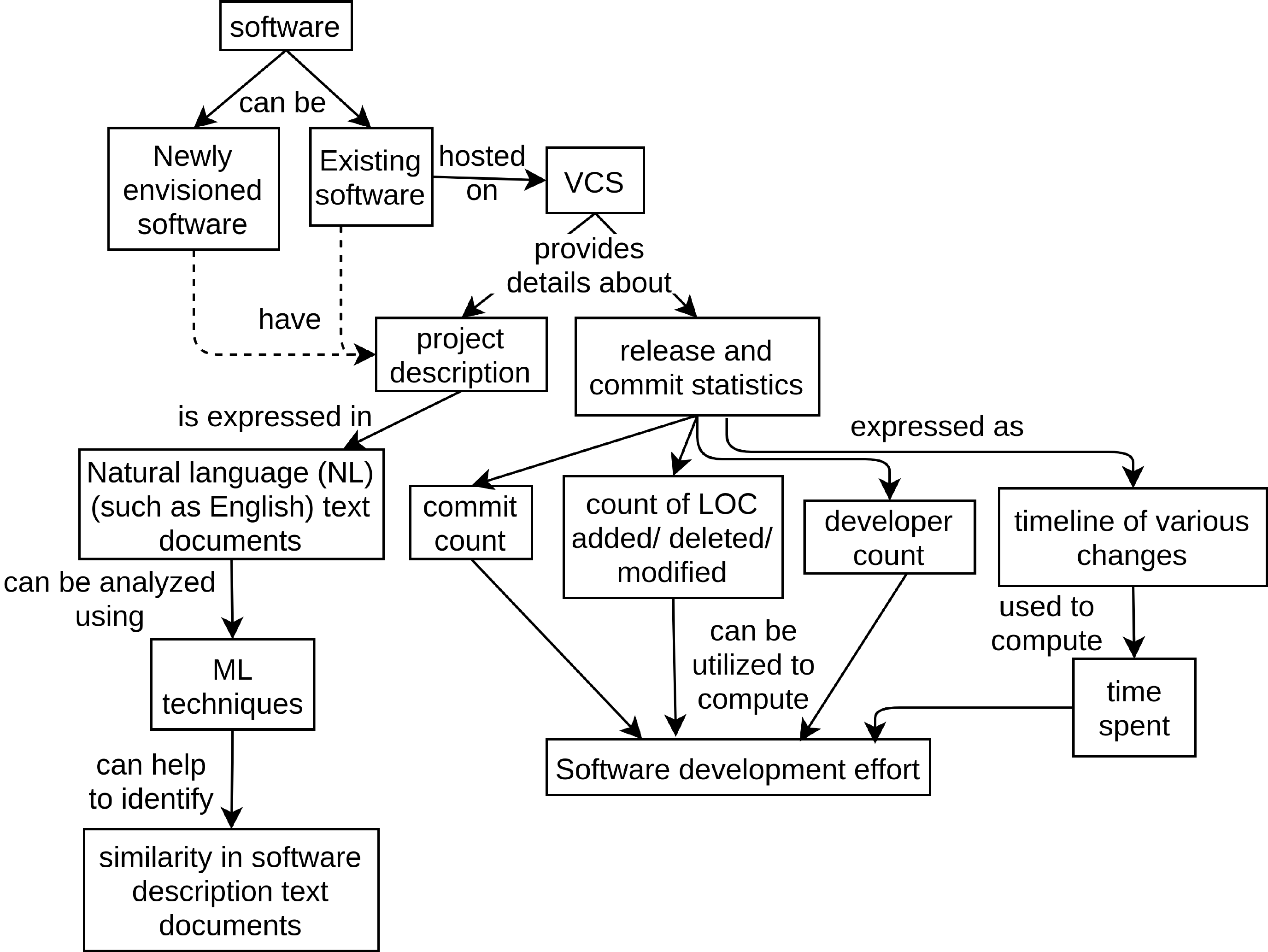}
\caption{Essential idea of our work}
\label{fig:basic-tenets}
\end{figure*}

\subsection{Essential ideas behind our system} 
\label{sec:basic-tenets}

Fig. \ref{fig:basic-tenets} depicts the essential ideas behind our system, listed as follows:
\begin{enumerate}
\item SDEE for newly-envisioned software is performed by considering the past software projects of similar nature  \cite{pressman2005software,sommerville2011software}. 

\item VCSs, such as GitHub and GitLab\footnote{\url{https://gitlab.com/explore}}, provide rich details about the actual software artifacts, the programmers who developed them, and the time spent in developing those artifacts. 
Some examples of high-level data available on various VCS hosts are the details of contributors working on a repository, the source code commits history, and the release timeline. 

\item The software requirements and the project description are expressed in a \emph{natural language (NL)} such as English. Today, ML techniques have matured to enable accurate and efficient detection of text document similarity. 
It is possible to identify existing software projects similar to those of a newly envisioned one.

\begin{definition}[Developer activity information]
\label{def:developer activity-information}
By \textbf{developer activity information of a software repository}, we refer to the following measures:
\begin{enumerate}
    \item \emph{Source code contribution by a developer:} The number of additions, deletions, and modifications of SLOC performed by a developer on a software.
    
    \item \emph{The timeline of various changes in a VCS:} It acts as a valuable metric for computing the time required to develop a software product. 
    
    \item \emph{The metadata information of the software repository:} The information such as the number of developers working on the software repository is essential for computing the effort required to develop the software product.
\end{enumerate}

A software developer is majorly involved in the following aspects of a software development project:
\begin{enumerate}
    \item Developing source code, which is added to the repository in the form of commits.
    \item \emph{Addressing the issues or tickets:} Changes to the source code are the most likely and frequent outcome in this case, and is reflected, ultimately, in the commits.
\end{enumerate}
While performing the effort computation, we only include the effort spent by the developers in \emph{code development}.

\textbf{Using developer activity information as SDEE metrics:}
\label{sec:intuition-developer-based}
\begin{enumerate}
    \item A significant part (more than 25\%) of the total software development effort is spent in source code creation  \cite{selamat2014new}.
    
    \item Developers  are the key contributors towards the source code creation and, thus, software development.
    
   \item A developer's contribution often depends on several factors, such as his productivity, participation, or activity  \cite{amor2006effort,gousios2008measuring}.

\end{enumerate}

Thus, leveraging the developer activity-based SDEE metrics as the first-class parameter in our SDEE model is expected to produce estimates with better accuracy.
\end{definition}

\item Developer activity information, such as the SLOC added, deleted or, modified by a developer, represents the most \emph{basic} unit of the software development effort contributed by a developer in developing a software project. 

\item Developer activity information present in VCS repositories can help provide a valuable measure of the effort spent developing it. Developer activity information can be derived from the release and commit statistics of various software projects tracked in the VCSs. Thus, to estimate the effort required to develop newly-envisioned software, the effort estimates derived using such information from the existing software projects with similar functionality can be leveraged.
\end{enumerate}



\subsection{Contributions}
\label{sec:contributions}

\begin{enumerate}
    \item We present a novel method to estimate the effort required to develop software. Given the software requirements of a newly-envisioned software, our system estimates the effort required to develop it. To derive the effort estimates, we use the developer activity information of the past software. 
    \item We propose new SDEE metrics based on the developer activity information. For instance, the the developer count (devCount) and time spent in developing a software project.
  \item By extracting the developer activity information for various software present on GitHub, we develop a dataset that can be leveraged to perform SDEE.
   \item By utilizing the product descriptions of software considered in our dataset, we develop a software similarity detection model $M$ using PVA. For a given description of software requirements, $M$ predicts the functionally similar software matches existing on GitHub.
 \item With the combination of the SDEE dataset and the trained software similarity detection model, we develop our SDEE tool for predicting the effort required to develop a newly-envisioned software, with a standardized accuracy (SA) value of 59.89\% over the random-guessing method.
 \item We present a comparison of our approach with the existing SDEE methods, and the results show that our model outperforms the ATLM  \cite{whigham2015baseline}, as it achieves a SA of 42.7\% when compared with it.
\end{enumerate}

\textbf{Paper organization:} Next, we discuss some of the related works and their shortcomings in Section \ref{sec:related-work}. Section \ref{sec:proposed-system}
highlights the critical design decisions and the implementation details of our system.
Section \ref{sec:performance-evaluation-and-comparison} gives the details of
the experiments conducted to verify the effectiveness of our system. Section \ref{sec:threats-to-validity} lists the threats to the validity of our approach. Finally, Section \ref{sec:conclusion} summarizes our work and suggests the future work.

\section{Background}
\label{sec:related-work}
\begin{figure*}
\centering
\includegraphics[width=120mm]{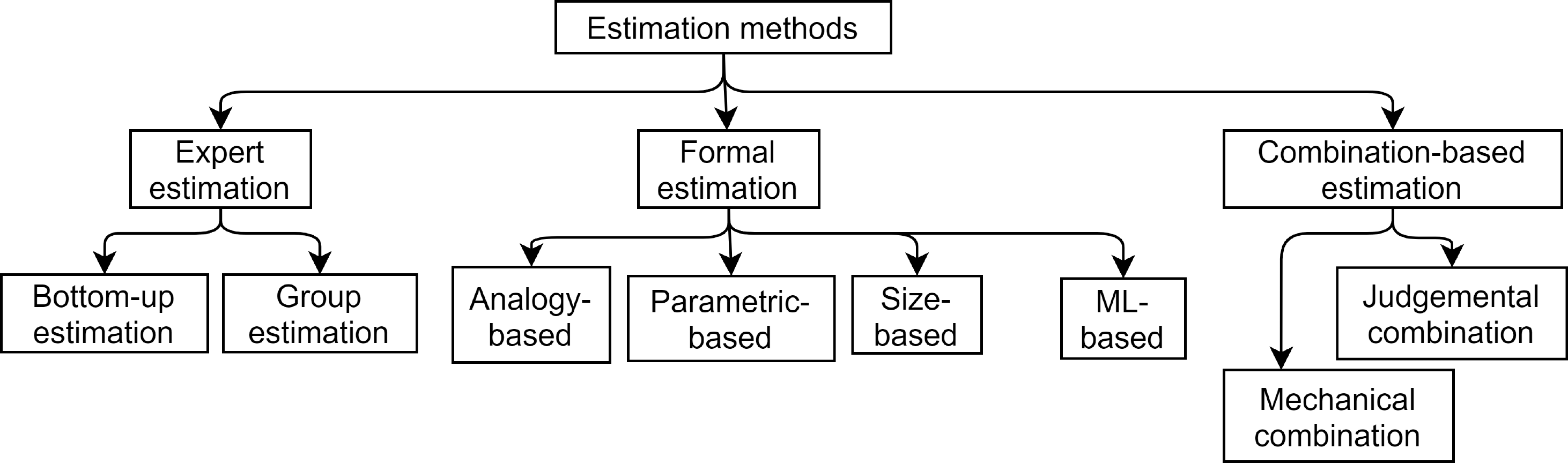}

\caption{Categorization of the existing effort estimation methods}
\label{fig:sdee-methods-hier}
\end{figure*}
\subsection{Existing approaches for SDEE}
\label{sec:existing-approaches}

Fig. \ref{fig:sdee-methods-hier} shows a broad categorization of the existing effort estimation techniques summarized by reviewing the existing literature \cite{molokken2003review}. Based on our literature review, we classify the existing SDEE techniques broadly into three major categories: \emph{Expert-based SDEE} methods, \emph{Formal SDEE} methods, and \emph{Hybrid SDEE} methods. A brief description of these methods is as follows:

 \begin{enumerate}
    \item \emph{Expert-based SDEE methods:} These are based on the judgments provided by the experts, such as the project managers, using the knowledge of software projects developed in the past  \cite{pressman2005software}. The expert-based SDEE methods are further sub-categorized as \emph{Bottom-up SDEE methods} and \emph{Group SDEE methods}. Bottom-up methods perform the SDEE by using the following procedure: a) Dividing the complete development process into sub-tasks, b) Estimating the effort required to perform these sub-tasks, c) Finally, summing these effort values to obtain an overall effort estimate. The estimates for performing the sub-tasks are obtained from the specific people in charge of developing them.
    On the contrary, Group SDEE methods comprise consulting a group of experts to derive the  effort estimate for developing a software project  \cite{wu1997comparison}. 
    
    \item \emph{Formal SDEE methods:}
    These are based on the software development effort estimates derived using a mathematical quantification step \cite{jorgensen2009software}, such as a formula derived from historical data from past projects. These can be further classified as \emph{Analogy-based SDEE methods}, \emph{Parametric SDEE methods},  \emph{ML-based SDEE methods}, and \emph{Size-based SDEE methods}. Analogy-based SDEE methods follow \textbf{identifying one or more existing software projects similar to the project being developed} and then deriving the effort estimate by analyzing similar software projects. Similarly, Parametric methods are based on  \textbf{identifying the variables influencing the effort required to develop a software project}. The general mathematical form of such models is represented as:
        
        \begin{equation}
            \label{eq:general-parametric-models}
            E = A + B(e_{\nu})^{C}
        \end{equation}
        In the above equation, $E$ represents the effort in person-months, $e_{\nu}$ represents the estimation variable (e.g., source lines of code (SLOC) or function points \cite{albrecht1979measuring,dumke2016cosmic}), and $A, B$, and $C$ represent the empirically derived constants the past software data. Some of the prominently used Parametric SDEE methods include COCOMO \cite{barry1981software}, Putnam's model \cite{putnam1978general}, and SEER-SEM \cite{galorath2007software}.
    
    Size-based SDEE methods form a subset of parametric methods, as these use size-based metrics, such as SLOC and function-points,  to compute the effort estimates for developing a software project. Some other examples of such metrics are use-case points (UCPs) \cite{winters1900applying}, story points \cite{cohn2005agile}, and object points \cite{minkiewicz1997measuring}. Recently, COSMIC Functional Size measurements (COSMIC FSM) are reported to overcome the shortcomings of LOC-based metrics, and outperform the existing FP-based methods \cite{di2016web,abualkishik2017study}. COSMIC FSM is based on the Functional User Requirements (FURs) of the software that are independent from any requirements or constraints about their implementation. They have been used for code size estimation of mobile applications \cite{d2015cosmic}, UML models \cite{de2020design}, Web applications \cite{di2016web,abrahao2018definition}, etc.   
    
    \item  \emph{Hybrid SDEE methods:} These methods use a combination of two or more existing effort estimation methods. Some of the prominent hybrid SDEE methods derived from the literature \cite{briand2002resource} are \emph{Mathematical Hybrid SDEE methods} and \emph{Judgemental Hybrid SDEE methods}. 
    Mathematical Hybrid SDEE methods perform the SDEE using the analogy-based methods and the bottom-up SDEE methods, whereas Judgemental Hybrid SDEE methods perform SDEE using the expert-based SDEE methods and the parametric SDEE methods.
\end{enumerate}
 
\subsection{SDEE methods based on developer activity information and ML techniques}

Some existing studies \cite{kocaguneli2011exploiting, menzies2017negative} present the variation of analogy-based SDEE methods where they use the k-Nearest Neighbor (kNN) method with the euclidean distances to detect the similar software matches from existing SDEE datasets, such as the PROMISE repository \cite{menzies2012promise}. A baseline of SDEE methods was established by \cite{whigham2015baseline} when they introduced their Automatically Transformed Linear Baseline model, which is defined as a linear regression model. The ML-based methods, such as Artificial Neural Networks (NeuralNet),  are also employed to train SDEE models \cite{minku2013software}. 

ATLM \cite{whigham2015baseline} and ABE \cite{kocaguneli2011exploiting} use  features such as the analyst's experience, programmer's capability, and application experience when trained using the existing datasets, such as the PROMISE repository. Similarly, the LOC \enquote{straw man} estimator \cite{menzies2017negative} uses the LOC of the software projects. Both ABE and the LOC \enquote{straw man} estimator use the weighted euclidean distance measure to determine the top-k similar matches. The weighted euclidean distance between two feature values x and y can be expressed as:

\begin{equation}
    \label{eq:euclidean distance}
    dist(x, y) = \sqrt{\sum_{i=1}^{n} w_i (x_i - y_i)^2}
\end{equation}

where $x_i, y_i$ are values normalized to [0,1], and $w_i$ is a feature-weighting factor (defaults to $w_i = 1$).

In contrast, ATLM has a linear regression form expressed as:
\begin{equation}
    \label{eq:ATLM}
    y_i = \beta_0 + \beta_1x_{1i} + \beta_2x_{2i} + ..\beta_nx_{ni} + \epsilon_i 
\end{equation}

where $y_i$ represents the quantitative response variables, the $x_i$ represents the  explanatory variables, and the $\beta_i$ is determined using a least-squares estimator \cite{neter1996applied}. Similarly, the study in \cite{minku2013software} is based on training Artificial Neural Networks (NeuralNet) using various datasets to predict the effort estimates.


An empirical approach for analyzing the developer activity traces present in various VCSs to categorize the software developers as \{occasional contributors, full-time developers\} is presented in \cite{robles2014estimating}. They present the analysis for data collected by Bitergia\footnote{\url{https://www.bitergia.com/}} for OpenStack Foundation\footnote{\url{https://https://www.openstack.org/foundation/}}. CVSAnalY is deployed to collect the analytics from the source code management repository. Developers of the project are then mailed a survey to inquire about the time devoted to developing the project. The obtained survey data is used to train a binary classification model by labeling the dataset entries as \{occasional contributors, full-time developers\}. The approach is claimed to propose a realistic estimation with a smaller weight (of the total effort) assigned to the more error-prone estimations.

Similarly, a Change Impact Size Estimation (CISE) approach for software development is introduced \cite{asl2013change}. The approach is designed by integrating the static and dynamic approaches mentioned in \cite{kama2011change}. Change Impact Analysis (CIA) is performed by tracing the affected requirements by change requests to the impacted software artifacts and codes before they are implemented. The essential requirements types are an addition, a modification, and a deletion. The CISE works by first finding the classes affected by the changes (by performing a CIA) and then performing estimation for change in size. The study claims CISE to be 93.7\% accurate.

A comprehensive study of the developer activity patterns in Free/Libre/OSS (FLOSS) projects is presented by clustering developers around different time-slots and days of a week \cite{capiluppi2013effort}. The study compares the developer activity patterns in a company-driven environment with those in a community-driven environment. The study is conducted by analyzing the developer activity patterns based on the number of commits, lines of code added, deleted, modified as recorded on several days of a week, and different hours. In our work, we utilize these counts, viz., the number of commits, lines of code added, deleted, and modified, extracted from various software present on GitHub to estimate the effort expended to develop the software.

\subsection{Computing the developer coding effort estimates for OSS}

\cite{asl2013change} introduce a Change Impact Size Estimation (CISE) approach for software development. The approach is designed by integrating the static and dynamic approaches mentioned in \cite{kama2011change}. Change Impact Analysis (CIA) is performed by tracing the affected requirements by change requests to the impacted software artifacts and codes before they are implemented. The major requirements types are an addition, a modification and a deletion. The CISE works by first finding the classes affected by the changes (by performing a CIA), and then performing estimation for change in size with an accuracy of 93.7\%. Similarly, there are existing works on developing defect prediction solutions by measuring code changes in terms of SLOCs additions, deletions, or modifications of software releases \cite{bell2011does,hassan2009predicting}. 

\cite{capiluppi2013effort} study the developer activity patterns in Free/Libre/OSS (FLOSS) projects by clustering developers around different time-slots and days of a week. The developer activity patterns in a company-driven environment are compared with those in a community-driven environment based on the number of commits, lines of code added, deleted, and modified as recorded on various days of a week, and at various hours of a day. Similarly, \cite{robles2014estimating} propose an empirical approach for analyzing the developer activity traces present in various VCSs to categorize the software developer as \{occasional contributors, full-time developers\}. They present the analysis for data collected by Bitergia\footnote{\url{https://www.bitergia.com/}} for OpenStack Foundation\footnote{\url{https://https://www.openstack.org/foundation/}}. CVSAnalY is deployed to collect the analytics from the source code management repository. Developers of the project are then mailed a survey to inquire about the time devoted to developing the project. The obtained survey data is used to train a binary classification model by labelling the dataset entries as \{occasional contributors, full-time developers\}. The approach is claimed to propose realistic estimation with a smaller weight (of the total effort) assigned to the more error-prone estimations.

\cite{di2017developer} stress that developer's personality characteristics and the manner in which they perform the code changes play a major role in software development. They study the developer personality characteristics by measuring \enquote{how focused the developers are while making code changes?} and \enquote{how scattered these changes are?} to devise a novel bug prediction model. The model is validated using 26 Apache software projects and comparing with four state-of-the-art prediction models based on size metrics and process metrics. Similarly, \cite{catolino2018enhancing} leverage the developer-related factors to enhance the existing change-prediction models by inspecting \enquote{how developers perform modifications?} and \enquote{how complex is the development process?} The idea is to use the complexity of the development process to identify the change-prone classes, and then focus the developer's attention to such classes. The authors perform an empirical analysis of three developer-based change prediction models considering the 408 releases, 193,274 commits, and 657 developers' information of 20 GitHub projects, and propose a novel hybrid change prediction model which exploits the developer-, process-, and product-metrics to detect the change-proneness of source code classes.

Similar to the existing works \cite{capiluppi2013effort,asl2013change,robles2014estimating,kocaguneli2011experiences}, we utilize the developer activity information such as the count of developers and time (in days) required to estimate the effort required to develop an existing GitHub software repository. We store these SDEE estimates as a dataset, which is then used by our SDEE tool to provide the effort estimates for newly-envisioned software.



\subsection{Developing a software description similarity detection model for detecting similar software matches}
 
 There exists a considerable amount of literature highlighting the use of various NL processing techniques to develop software description similarity detection models for various recommendation tasks, such as performing recommendation of API methods \cite{thung2013automatic}, third-party libraries \cite{chen2016similartech}, and research papers \cite{beel2013introducing}. Two prominent techniques used in the existing works are \emph{Term-frequency Inverse Document Frequency (TF-IDF)} \cite{salton1988term} model and \emph{Word2vec model} \cite{mikolov2013distributed,mikolov2013linguistic}.
 
 
 TF-IDF methods compute document similarity based on the term-frequencies for the terms or words present in a document, while the Word2vec model works by learning the word representations for the words present in an input document, to predict the co-occurring words in the same sentence. 

\cite{trstenjak2014knn} present the use of K-Nearest Neighbour (KNN) algorithm with TF-IDF method for text-classification of documents belonging to four different categories, viz., sports, politics, finance, and daily news. The results of the experiments show that the classification algorithm is sensitive to type-of-documents. The classification algorithm yields the highest accuracy of 92\% in the case of documents belonging to the \emph{sports} category, while the lowest as 65\% in the case of the \emph{daily news} category.

 We use the Paragraph vectors algorithm (PVA) \cite{le2014distributed} to develop our software description similarity detection model. The PVA is inspired by the Word2vec method and is also termed as a vector, as it generates fixed-length feature representations (or vectors) corresponding to the text documents provided as input. 

\textbf{Limitations of the TF-IDF method:}

Since the TF-IDF method is based on the bag-of-words approach, some of its limitations are \cite{ramos2003using}:
\begin{enumerate}
    \item \label{lim:one} It is unable to capture the semantic information of the text on which it is trained. 
    \item \label{lim:two} TF-IDF method does not consider the ordering of different words in a text document or the co-occurrences of words in different documents.
    \item TF-IDF might take much time to train on a large vocabulary.
\end{enumerate}

Due to the limitations specified in point \ref{lim:one} and \ref{lim:two} above, TF-IDF is not able to capture the documents that differ in words having the same meaning i.e. synonyms. For instance, happy, content, and joyful will be treated as different words by TF-IDF method. Further, the method would also label two documents as different, if they convey the same idea using different descriptions. For instance, \emph{it was a busy day}, and \emph{I was occupied with a lot of work today} both convey the same meaning, but will be labeled as different. Similarly, the TF-IDF method would label two project descriptions as different, if they use different words to convey the same meaning. For instance, consider the following software requirements:
\begin{enumerate}
    \item \emph{Require a fast, accurate, and anti-obfuscation library detection tool}, and
    \item \emph{Require a light-weight, effective, and obfuscation-resilient tool.} 
\end{enumerate}

Both the examples convey the same software requirements, but with different use of words, and hence are labeled as different. Thus, these limitations make TF-IDF an inappropriate measure for detecting similarity in our case, i.e., when dealing with software requirements and software project descriptions.

\begin{figure*}
\centering
\includegraphics[width=0.9\columnwidth]{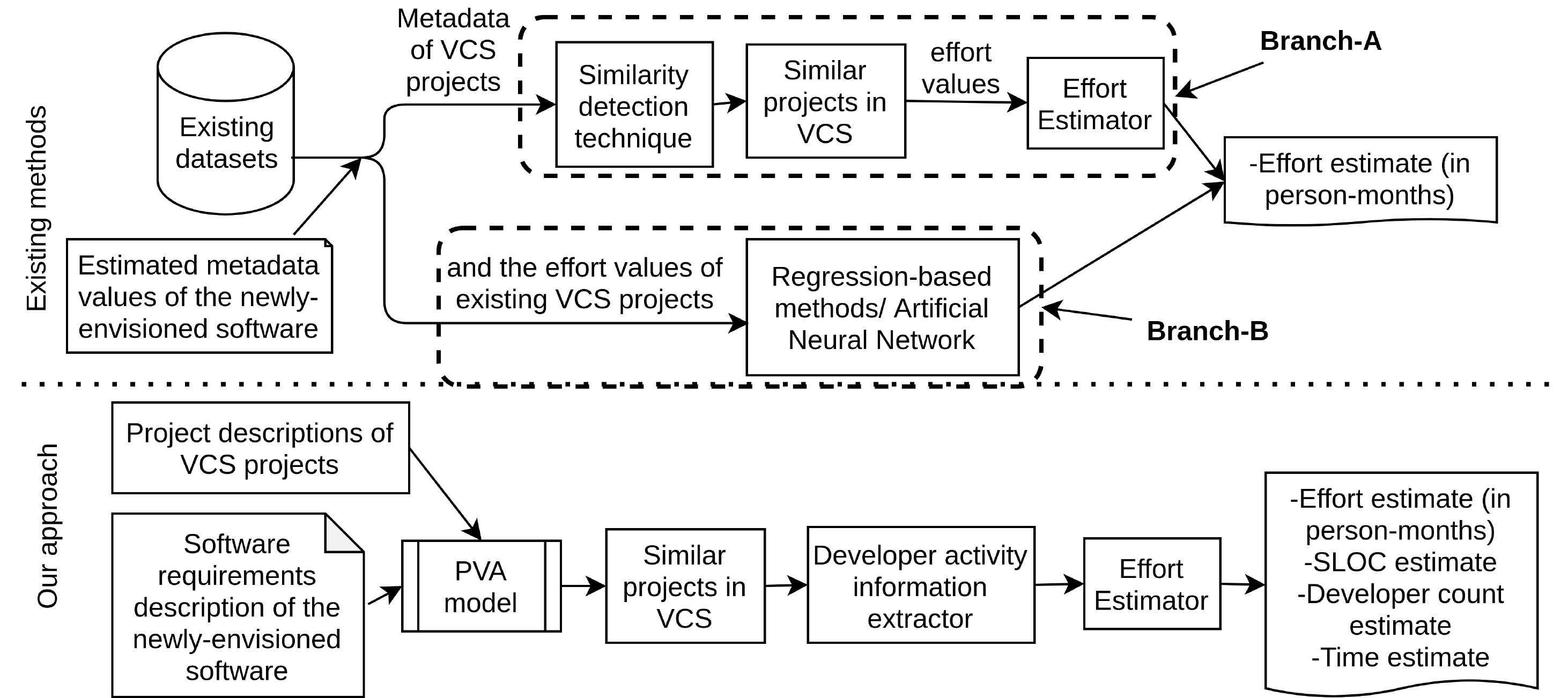} 
\caption{High-level comparison at approach-level}
\label{fig:high-level-comp}
\end{figure*}

\subsection{High-level differences of our approach from the existing methods}
\label{sec:high-level-sdee-comp}
Fig. \ref{fig:high-level-comp} presents a broad idea of how our approach differs from the existing SDEE methods. The existing effort estimation methods, such as the Analogy-based effort estimation (ABE) \cite{kocaguneli2011exploiting} and LOC \enquote{straw man} \cite{menzies2017negative}, generally follow the branch (a) approach shown in the figure, where, to estimate the effort for a newly-envisioned software $z$, the effort values of software projects developed in the past and similar to $z$ are used. On the contrary, the regression-based methods, such as the Automatically Transformed Linear Baseline Model (ATLM) \cite{whigham2015baseline} and Artificial Neural Network (NeuralNet) \cite{minku2013software} directly learn and estimate the effort value of $z$ using the effort values of the past software projects, represented by branch (b) of the figure. The existing SDEE methods mostly train the models on the metadata values of the software projects developed in the past, present in the existing effort estimation datasets, such as the PROMISE repository \cite{boetticher2007promise}.

In our approach, we hypothesize that \enquote{more relevant similar-software matches can be detected if we use the software project descriptions as the input for detecting similar software projects.} Also, the developer activity information has been claimed to be an accurate representation of the actual development effort on a software project  \cite{capiluppi2013effort, robles2014estimating}. Therefore, we use the software project descriptions of GitHub repositories to determine the similar software projects (to $z$) and use the developer activity information of the similar software matches to compute the effort estimate for $z$.


\begin{table}
\caption{Table of Notations}
\centering 
    \begin{tabular}{r c p{12cm}} 
    \toprule
    
    $R$ & $\triangleq$ & The set of OSS repositories present on GitHub considered by us.\\

     $e^{r}$ & $\triangleq$ & Effort required to develop $r \in R$, computed using equation \ref{eq:software-development-effort}\\
    
    $D^{r}$ & $\triangleq$ & The set of developers involded in developing $r \in R$.\\
     
      $t^{r}$ & $\triangleq$ & The total amount of time spent in developing $r \in R$.\\

    

       $C$ & $\triangleq$ & The set of different software categories specified on MavenCentral\footnote{\url{https://mvnrepository.com/}}.\\

    $H$ & $\triangleq$ & The set of software project descriptions of $R$.\\
           
       $P$ & $\triangleq$ & The set of PVA input parameter variation scenarios for training the software similarity detection models.\\
    
    $M$ & $\triangleq$ & The set of software similarity detection models trained using PVA on $H$.\\

     $Y$ & $\triangleq$ & The test-bed developed to test the performance of software similarity detection models.\\
    
    $\phi$ & $\triangleq$ & The vector representation of a project description.\\
    
    $\alpha$ & $\triangleq$ & The cosine similarity score between two PVA vectors.\\
    
   $\hat{\alpha}$ & $\triangleq$ & The threshold of cosine similarity between two PVA vectors to categorize them as similar.\\
    
    $\psi$ & $\triangleq$ & Number of training samples used for training a software similarity detection model. \\
    
    $\gamma$ & $\triangleq$ & It refers to the dimensionality of the feature vectors used to develop the dataset, referred to as the PVA vector size. \\
    
    $\beta$ & $\triangleq$ & Number of training iterations or epochs used for training a software similarity detection model.\\

    \bottomrule
    \end{tabular}
\label{tab:TableOfNotation}
\end{table}

To the best of our knowledge, such use of software project descriptions and the developer activity information has not been done in performing the SDEE task. However, since we have chosen the OSS (GitHub) repositories  to develop our dataset, our SDEE model provides estimates for only the software developed in an open-source environment. This is because the software development for open-source software is generally performed by source code contributors who do not have fixed working hours or work patterns, whereas the proprietary software is developed by full-time working employees  \cite{capiluppi2013effort} having fixed working patterns. Thus, the developer activity might differ due to the different working patterns. 

\begin{figure*}
\centering
\includegraphics[width=\linewidth]{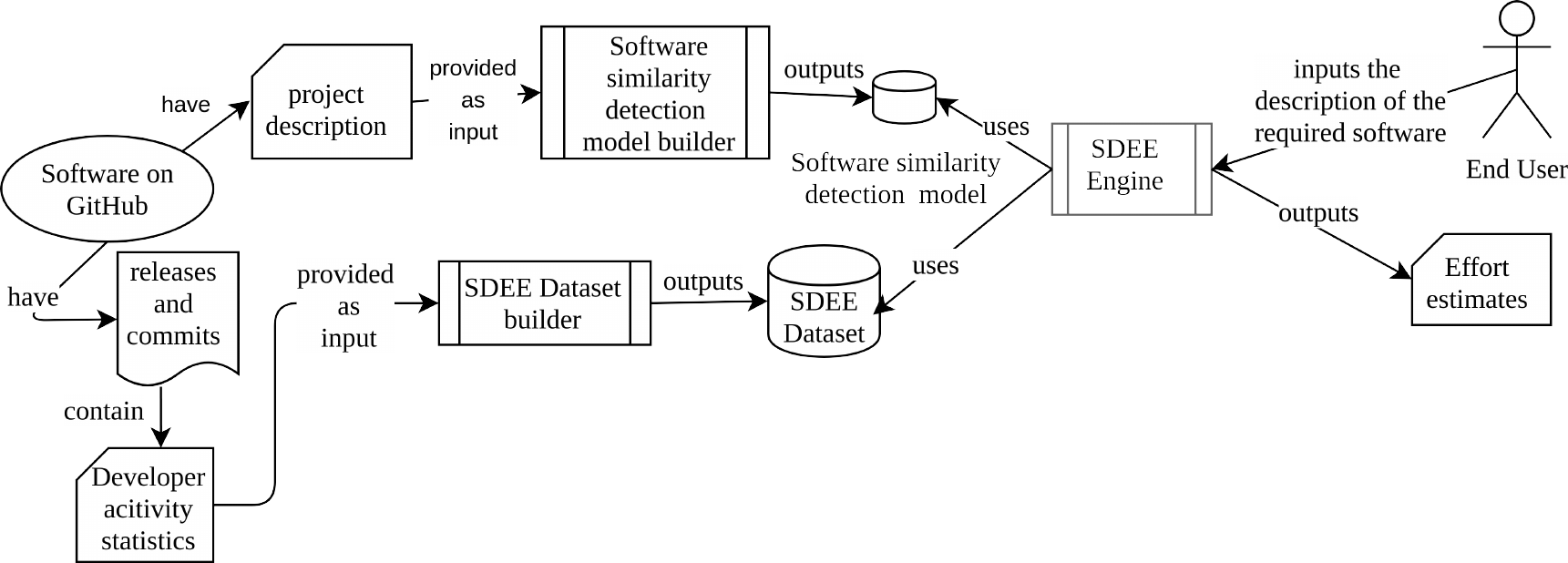}
\caption{Details of the proposed system - a system-context diagram}
\label{fig:basic-approach}
\end{figure*}

\section{Proposed system}
\label{sec:proposed-system}

The main objective of our approach is: 

\emph{Given the requirements description of a software project, estimate the effort required to develop it.}

To achieve this objective, we develop three major software artifacts (shown in Fig. \ref{fig:basic-approach}):
\begin{enumerate}
    \item A novel \emph{SDEE dataset} based on the developer activity information of software present on GitHub,
    \item A \emph{software similarity detection model} (developed using PVA) trained on the project description documents of software considered to develop the above dataset,
    \item An \emph{SDEE tool} that utilizes the SDEE dataset and the software similarity detection model to provide effort estimates. 
\end{enumerate}

Table \ref{tab:TableOfNotation} shows the notation used for various terms in this paper.

\subsection{Developer activity-based SDEE metrics}
\label{sec:SDEE-metrics}
An OSS development broadly comprises developers contributing source code in an OSS repository. To measure the software development activity quantitatively, we define novel SDEE metrics for an OSS repository $r$, as follows:
\begin{enumerate}
 


\item $|D^r|$: The total number of developers involved in developing $r$. It is measured in \emph{persons}.

\item $T^r$: The total amount of time spent in developing $r$, mathematically represented as:
\begin{equation}
\label{eq:software-development-time}
t^r \mbox{=}\ \sum_{i=1}^j (t_s^i - t_e^i)/j
\end{equation}
where $t_s^i$ represents the start-time of developing the ith release of repository r, and $t_e^i$ represents the end-time of the ith release. It is expressed in days, months, or years.    

\item $e^r$: The effort expended to develop $r$, mathematically represented as:

\begin{equation}
\label{eq:software-development-effort}
e^r \mbox{=}\ |D^r| * t^r
\end{equation}

It is measured in person-months, or person-years, or person-days, depending upon the units of $|D^r|$ and $t^r$. 

\end{enumerate}

\subsection{Steps of our approach}
\label{sec:steps-of-approach}
The pivotal steps in our approach with the relevant design decisions addressed at each step are listed below:   

\begin{enumerate}
    \item \textbf{Processing the GitHub software repositories and developing the SDEE dataset}
    \begin{enumerate}
        \item Download the software categorization listed on MavenCentral.
        
        \emph{Design Decision: Why choose the software categorization from MavenCentral?}
        \item Download \label{step:downloading-project-desc} a collection of GitHub software repositories belonging to some of the significant software categories. We used the ones specified by MavenCentral. Also, extract the developer activity information and download the software project descriptions of the respective GitHub repositories.
        \item Compute the effort estimates for developing the considered software repositories by utilizing their developer activity information. The critical design decisions considered in this step were:  
        \begin{enumerate}
            \item \emph{How do we ensure our dataset's homogeneity in terms of the breadth of software types or domains?}
            \item \emph{Why is only GitHub chosen to download software repositories? How do we select a software repository?}
            \item \emph{What developer activity information should we collect from the selected repositories? How can we determine whether the collected information helps estimate the software development effort?}
            \item \emph{How do we utilize the developer activity information to compute the software development effort estimates?}
        \end{enumerate}
    \end{enumerate}
    \item \textbf{Developing the software similarity detection model and obtaining the PVA vectors}
    \begin{enumerate}
        \item Train a software similarity detection model using the PVA on the software project descriptions downloaded in Step \ref{step:downloading-project-desc}. The critical design decisions considered in this step were:
        
        \begin{enumerate}
            \item \emph{Why use PVA and how to choose its tuning parameters?}
            \item \emph{Why do we use vector representations of software product descriptions?}
        \end{enumerate}

        \item Obtain the PVA vectors for each of the product descriptions downloaded in Step \ref{step:downloading-project-desc}. 
        
            \emph{Design Decision: What is the minimum threshold value of  $\alpha$ between two PVA vectors  that indicate a significant similarity between the corresponding software project description samples?}
    \end{enumerate}
    \item \textbf{Developing the SDEE tool to perform the SDEE for a newly-envisioned software $z$}
    \begin{enumerate}
        \item Design a suitable GUI for the interaction with the end-user. The salient design decisions considered at this step were:
        \begin{enumerate}
            \item \emph{What is the input format for specifying the requirements description of software for which the development effort is predicted?}
            \item \emph{What type of details are required to be specified in the requirements description provided as input to the tool?}
        \end{enumerate}

        \item Compute the effort estimate for developing $z$ by leveraging the SDEE dataset and the PVA vectors.
        \begin{enumerate}
            \item Obtain a PVA vector $\phi'$ for the requirements description submitted by the end-user. 
            \item Fetch \label{step:fetch-sim-vecs} all the PVA vectors $\phi \in V$, such that the similarity $\alpha$ between $\phi$ and $\phi'$ is above $\hat{\alpha}$.
            \item Derive an effort estimate for $z$ by utilizing the similar software matches' effort values fetched in Step \ref{step:fetch-sim-vecs}.
            
            \emph{Design Decision: How do we compute the final effort estimate by utilizing the similar software matches' effort values?}
        \end{enumerate}
    \end{enumerate}
\end{enumerate}

In the following sections, we describe developing our software artifacts and the rationale for design decisions addressed in this process.

\subsection{Design considerations in our approach}
\label{sec:design-details}
In this section, we discuss the rationale behind design decisions taken by us while implementing our approach.

\begin{enumerate}
    \item \textbf{Obtain the software categorization:} \label{step:select-soft-categorization}
    To ensure our dataset's homogeneity in terms of the software types, it is necessary first to identify what types of software constitute the major portion of professional software. MavenCentral is one such source of software categorization taxonomy. Thus, we first fetch the software categorization listed by Maven (MVN) Repository\footnote{\url{https://mvnrepository.com/}}, then select the software projects under each of those categories to develop the dataset. The MVN repository portal contains around 150 software category web pages, with each category-specific web page containing the description and web links of the categorized software. These categories are utilized as topics to fetch the GitHub links of software using the GitHub REST API\footnote{\url{https://developer.github.com/v3/}}.
    
    \item \textbf{The rationale for choosing GitHub:} Though several OSS hosts such as SourceForge.net\footnote{\url{https://sourceforge.net/}}, GitHub, GitLab, and BitBucket\footnote{\url{https://bitbucket.org/dashboard/overview}}, only a few offer support for extracting the developer activity information.  We selected GitHub over the other OSS hosts for the following reasons:
\begin{enumerate}
\item SourceForge.net does not provide the details of source code development for the software repositories hosted by it. Hence, no developer activity information statistics were available.
\item GitLab and BitBucket, the newer ones, have very few OSS repositories that match our selection criteria. For instance, GitLab shows only 12 OSS repositories with $>$500 stars\footnote{\url{https://gitlab.com/explore/projects/starred}}. \end{enumerate}

    \item \textbf{Repository selection rationale:} We select the top 100 repositories provided by the GitHub RESTful API for each input category. The additional constraints \cite{coelho2020github,han2019characterization} applied to filter the OSS repositories under each of the selected software categories (selected in Step \ref{step:select-soft-categorization} above) are: 
\label{pt:url-extraction-constraints}
\begin{enumerate}
\item \emph{Size Constraint:} We select the repositories with a size of more than $5$ MB. The 5 MBs threshold was taken since we observed that the most non-trivial OSS repositories were larger than 5 MBs in their source code content. This means that the repositories with a size less than 5 MBs had minimal SLOC count.   
\item \emph{Date Constraint:} To ensure a repository is not very old or inactive, we add a date constraint to our search. We select the repositories which have been updated at least once in the last three years.
\item \emph{Stars Constraint:} We applied this constraint to ensure that the repositories selected are well-liked to some extent and hence being used by a certain number of developers. We selected the repositories with more than $500$ stars. We sorted our search results based on the stars earned by them and selected the top $100$ results under each category. 
\item \emph{Keyword constraint:} To filter the software repositories belonging to a specific category, we perform a GitHub REST API query search with the category names as the input keywords. 
\end{enumerate}  

\item \textbf{Utilizing developer activity information to compute effort estimates:} \label{step:developer activity-information}
The usefulness of developer activity information (Definition \ref{def:developer activity-information}) in performing the SDEE tasks  has been established by  \cite{robles2014estimating,capiluppi2013effort,gousios2008measuring,amor2006effort,rahman2013and,icse2014keynote}. To obtain the effort estimates for the considered software repositories, we define the SDEE metrics based on the repository's developer activity information. We obtain the software repositories' effort values by inserting the SDEE metrics' values in equation \ref{eq:software-development-effort}. Here, we consider the developer activity performed within the time-period $\langle$ start-date of release, date of the last commit performed on the release$\rangle$ to compute the release-wise effort estimates and store them in the table \emph{release\_effort\_estimate}. These release-wise estimates are then averaged to estimate the effort values of the software project, which are later utilized while performing the SDEE.

\item \textbf{The rationale for choosing PVA, its tuning parameters, and the similarity threshold $\hat{\alpha}$:} \label{sec:PVA}
The following are the main reasons for PVA's choice:  i) It allows us to compute the same length vectors for representing the software project descriptions. Keeping the length of such vectors the same for every source code sample is critical for implementing an efficient and fast system. ii) PVA outperforms the existing text representation methods and thus serves as an effective method of detecting text similarity  \cite{le2014distributed}.  Thus, we chose the PVA to develop our \emph{software similarity detection model} and compute the vector representations of our approach's software product descriptions.  The complete details of the parameter tuning of PVA are discussed in detail in Appendix-\ref{ap:PVA-parameter-tuning}.

\item \textbf{Computing PVA vectors for software product descriptions:}
\label{sec:computing-PVA-vectors}
To expedite the software similarity detection process, we perform the following steps:
\begin{enumerate}
    \item \label{step:a} Obtain the PVA vector representations of the software product descriptions using a suitable software similarity detection model (developed using Algorithm \ref{alg:training-pv-model}) and store them in the database. 
    \item Compute the cosine similarity ($cos\_sim$) of the PVA vectors for a reference vector of the same dimensions and store it in the database. The reference vector is randomly chosen and kept fixed for the dataset.
\end{enumerate}

\begin{figure*}
\centering
\begin{subfigure}{0.7\textwidth}
 \centering
  \includegraphics[width=\linewidth]{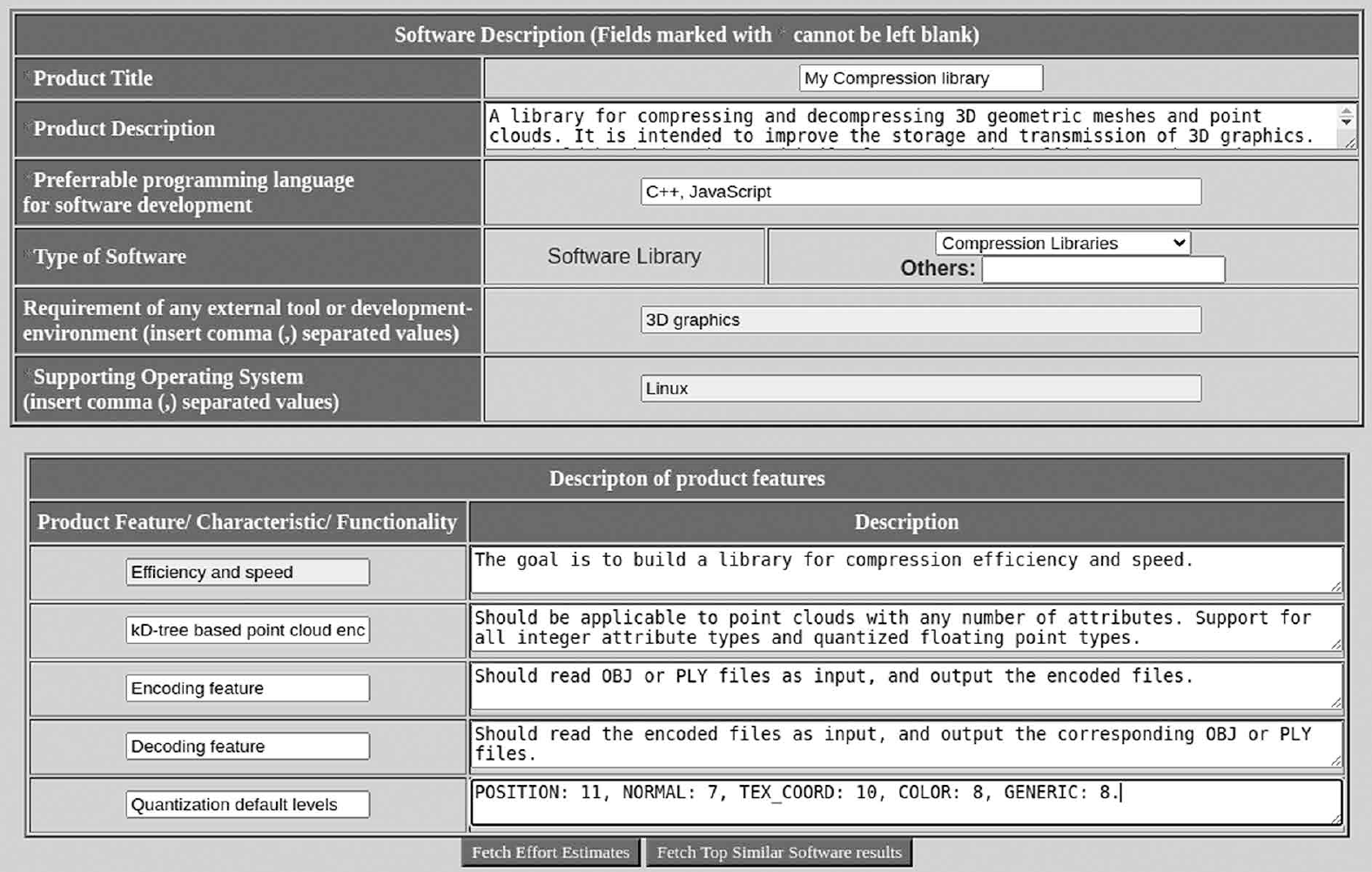} 
  \caption{Details submission page}
  \label{fig:submission-page}
\end{subfigure}%

\begin{subfigure}{0.63\textwidth}
\centering  
\includegraphics[height=.45\linewidth]{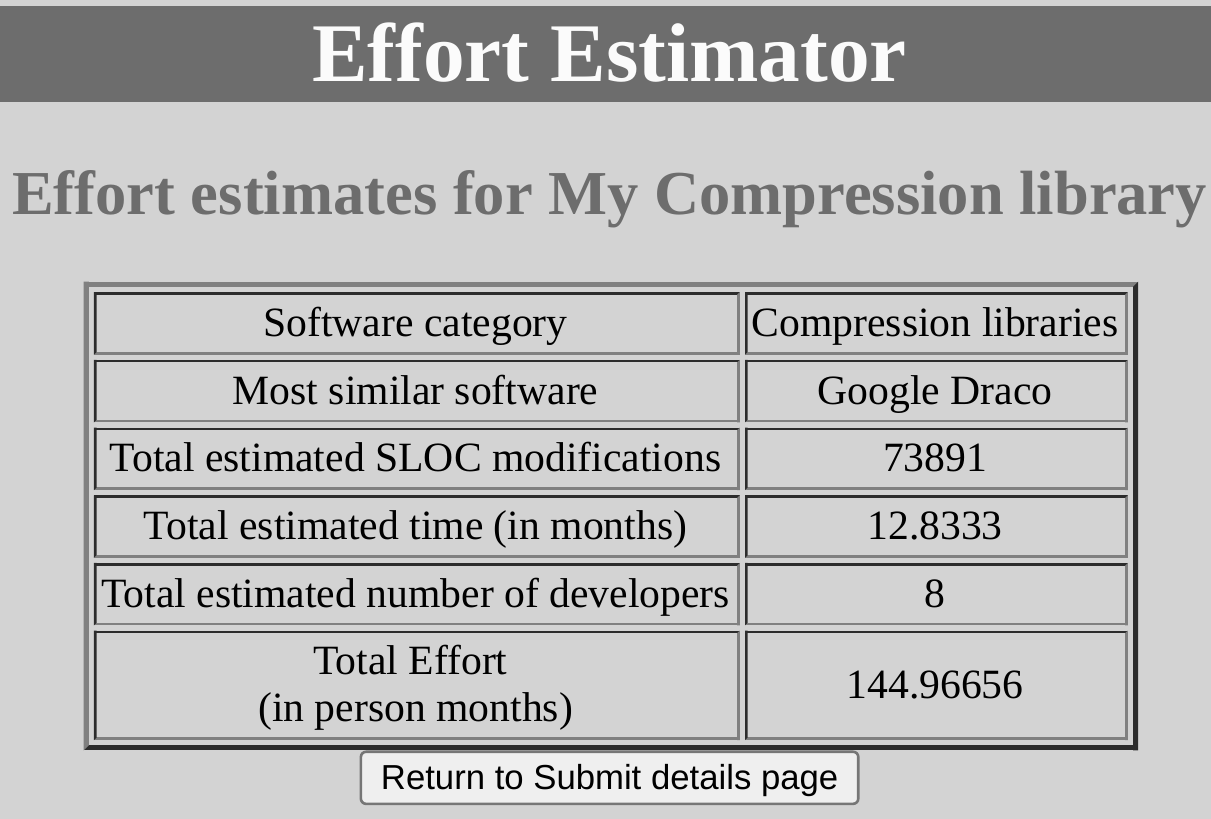} 
\caption{SDEE estimates for the input software details} 
\label{fig:tool-effort-estimate}

\end{subfigure}




\setlength{\belowcaptionskip}{-5pt}
\caption{SDEE tool interface}
\label{fig:sdee-tool-interface}
\end{figure*}

\item \textbf{Structure of inputs provided to the tool:}
\label{sec:tool-inputs}
Fig.  \ref{fig:sdee-tool-interface} shows the input and output pages of our tool's GUI. The software similarity detection model is likely to detect similar software with better accuracy if provided the input using the familiar format and structure as the training data. For instance, the software descriptions provided as input should describe the software's functionality to be developed, the features required to be developed in it, and a detailed description of all such functional requirements and software constraints.  Therefore, the format of the input expected by our tool is such that it captures the software's key functionalities and characteristics whose effort is to be estimated. For the newly-envisioned software, the end-user needs to enter the following inputs into the tool's GUI (shown in \ref{fig:submission-page}): 
\begin{enumerate}
    \item \emph{Software product title:} A suitable title for the software product.
    \item \emph{Software Description:} A brief description of the software product comprising its intended use, key functionalities, and characteristics.
    \item \emph{Preferred programming language for software development:} The programming language(s)  used for software development. In the case of multiple programming languages, the inputs can be entered in a comma-supported format. 
    \item \emph{Type of software:} The end-user must select a plausible software category and a sub-category from the respective drop-down lists. We extract 11 \emph{abstract} software categories from 150 software categories specified by MavenCentral, viz., \emph{i) Software library, ii) Software utilities \& plugin, iii) Software tool, iv) Software metrics, v) Software driving engine, vi) A software framework, vii) Software middleware, viii) Software client, ix) Software server, x) Software driver, and xi) Software file system.} 
    \item \emph{Supported Operating system:} In this input, the user specifies the operating system(s) the software product must support.
    \item \emph{Product features and feature description fields:} The end-user specifies the major software features and their description using these input fields.
\end{enumerate}

\begin{tcolorbox}
    
    \textbf{The rationale for deriving the abstract software categories:}
    To determine these abstract categories, we club the category names under their generic category types. For instance, SSH libraries, DNS libraries, JWT libraries can all be clubbed under the 
    \emph{Software library} category. On selecting one of the software categories, the end-user is displayed the specific sub-categories fetched from MavenCentral.
     \end{tcolorbox}

\textbf{Discussion on the similar software projects filtered by PVA:} As shown in Fig. \ref{fig:tool-effort-estimate}, for the given description of a compression library which is efficient and quick (in Fig. \ref{fig:submission-page}), we get the top matching result as \enquote{Google Draco} library\footnote{\url{https://github.com/google/draco}}. The next close match as projected by our tool is of \enquote{Centaurean Density} library\footnote{\url{https://github.com/k0dai/density}}. Both the libraries are intended to perform high-speed compression in an efficient manner as described in their GitHub project descriptions. This is because the tool identifies the best matching software by matching the input requirement description (shown in Fig \ref{fig:submission-page}) with the GitHub software project descriptions present in our database. 

\begin{figure*}
\centering
  \includegraphics[width=.4\linewidth]{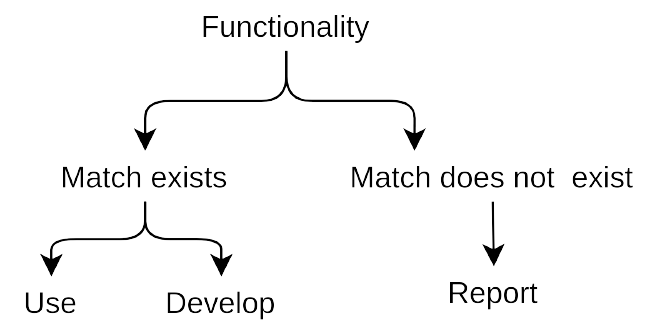}
  \caption{Detecting similar matches to a functionality}
  \label{fig:detect-sim-match}
\end{figure*}%

\item \textbf{Computing the effort estimate for developing a newly-envisioned software:}
\label{sec:computing-comb-effort-estimate}
Most of the GitHub hosted software repositories are used as software modules in developing more complex software systems. For instance, GitHub hosts a variety\footnote{For example, see this collection of projects \url{https://github.com/collections}} of software modules, such as third-party libraries, application development frameworks, and domain-specific applications. Since our dataset contains the effort values of such software, which act as modules in a more extensive software system, our tool can perform the SDEE for such software modules. Fig. \ref{fig:detect-sim-match} shows a decision tree for detecting existing software that matches the input.
The detailed procedure for estimating the software development effort of a newly-envisioned software $z$ is provided in Algorithm \ref{alg:estimating-effort-for-new-software}. To obtain the effort estimate for $z$, the effort values of the top-k similar software matches are combined using the Walkerden's triangle function  \cite{walkerden1999empirical}, defined as follows for k=3:

If the effort estimates of the first, second, and third closest neighbors are a, b, and c, respectively, then the effort required to develop $z$ is expressed as:
        \begin{equation}
            \label{eq:traingle-function}
Effort(z) \mbox{=}\ (3a + 2b + 1c)/6 
        \end{equation}
        
We use Walkerden's triangle function to obtain the effort estimates as it provides a weighted average result with the weights assigned based on the closeness of the top-k neighbors and results in smaller residual values than the non-weighted average method, i.e. when all top-k neighbors are assigned equal weights. We  achieved the best performance results for our SDEE method for k=2 (shown in Section \ref{sec:performance-evaluation-and-comparison}) and used it to compute the estimates.

\begin{algorithm}
\caption{Procedure to perform SDEE for a newly-envisioned software $z$}
  \label{alg:estimating-effort-for-new-software}
\begin{algorithmic}[1]
\REQUIRE $\hat{M} := $ The best-performing software similarity detection model developed in Algorithm \ref{alg:training-pv-model}.\\
$SDEE\_DB := $ Dataset containing the effort estimates computed using equation \ref{eq:software-development-effort} and the PVA vectors.\\
$\hat{\alpha} := $ The threshold cosine similarity score for PVA vectors.\\ 
$p_{z} := $ Project description of $z$.

\ENSURE $e_{z} := $ Effort estimate for the $z$. 

\STATE{$\phi_{d} := InferVector(p_{z}, \hat{M})$}
\STATE{$\phi_{ref} := FetchReferenceVector(SDEE\_{DB})$}
 \STATE {$\tau_{z} := ComputeCosineSimilarity(\phi_{z}, \phi_{ref})$}
 \STATE{$K_{z} := FetchTopKSimilarProjects(SDEE\_{DB}, \phi_{z}, \hat{\alpha})$}
\STATE{$E := FetchEffortValues(K_{z}, SDEE\_DB)$}
\STATE{$e_z := $ Return the averaged effort value using the Walkerden's triangle function equation \ref{eq:traingle-function} and the effort values $e \in E$.}
 \end{algorithmic}
\end{algorithm}
\end{enumerate}

\subsection{Implementation details}
\label{sec:impl-details}
We developed our software artifacts using the Python programming language due to the available expertise. The two significant artifacts driving our approach are:

\begin{figure*}
\centering
\includegraphics[width=0.7\columnwidth]{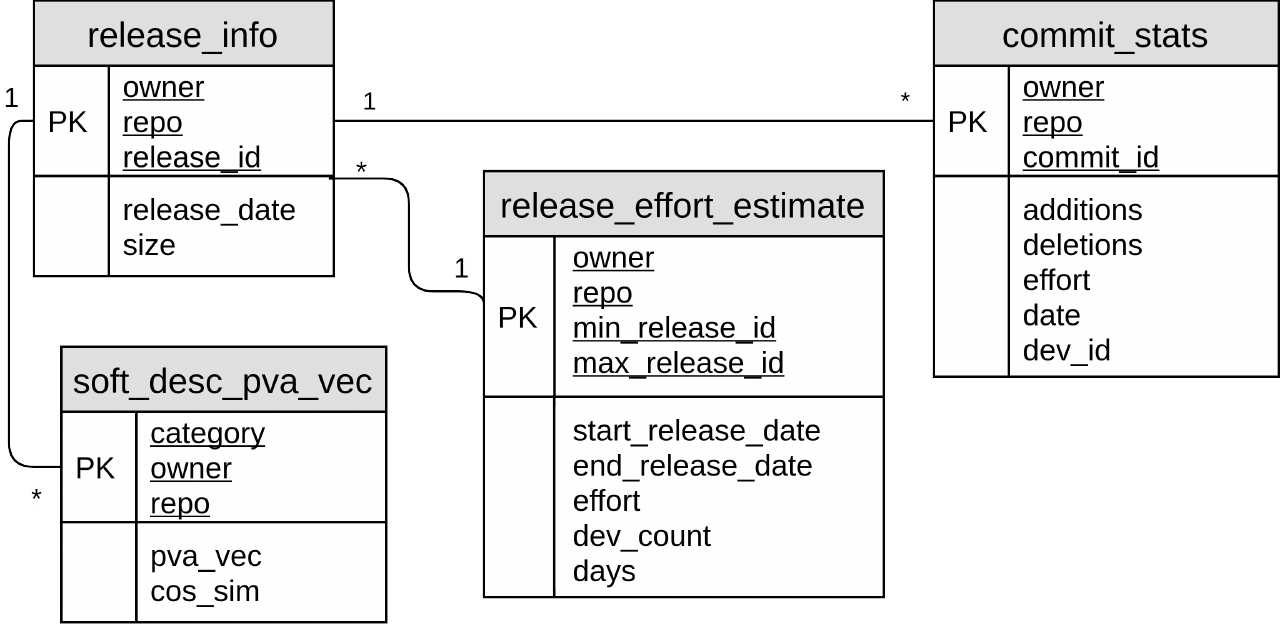}
\caption{Relational schema of the SDEE dataset}
\label{fig:SDEE-dataset-design}
\end{figure*}

\begin{enumerate}
    \item \textbf{SDEE dataset:} \label{step:developing-the-dataset} Fig. \ref{fig:SDEE-dataset-design} shows the relational schema of our SDEE dataset. The dataset mainly comprises the following: 
    \begin{enumerate}
        \item Information about the releases and commits in various VCS repositories. 
        \item Software development effort estimates computed by inserting the developer activity-based SDEE metrics values in equation \ref{eq:software-development-effort}.

      \item Metadata information, such as the developers involved in developing various software repositories.

        \item PVA vector representations of the \emph{software product descriptions} extracted from VCS repositories.
    \end{enumerate}

    The dataset contains the above-listed information extracted from $\approx$ 13,000 GitHub repositories belonging to 150 different categories of software.  The SDEE metrics values' extracted from various VCS repositories and the associated software estimates are retained as effort estimates in the SDEE dataset. The dataset is used by the SDEE tool to compute the effort estimates for developing the newly-envisioned software. It is easy to add data from additional sources. The details of the relational tables are described in Appendix \ref{ap:SDEE-dataset-details}.
  
    Our SDEE dataset and tool are available at \url{https://doi.org/10.21227/d6qp-2n13} and \url{https://doi.org/10.5281/zenodo.5095723}, respectively.

\begin{figure*}
\centering
\includegraphics[width=0.8\columnwidth]{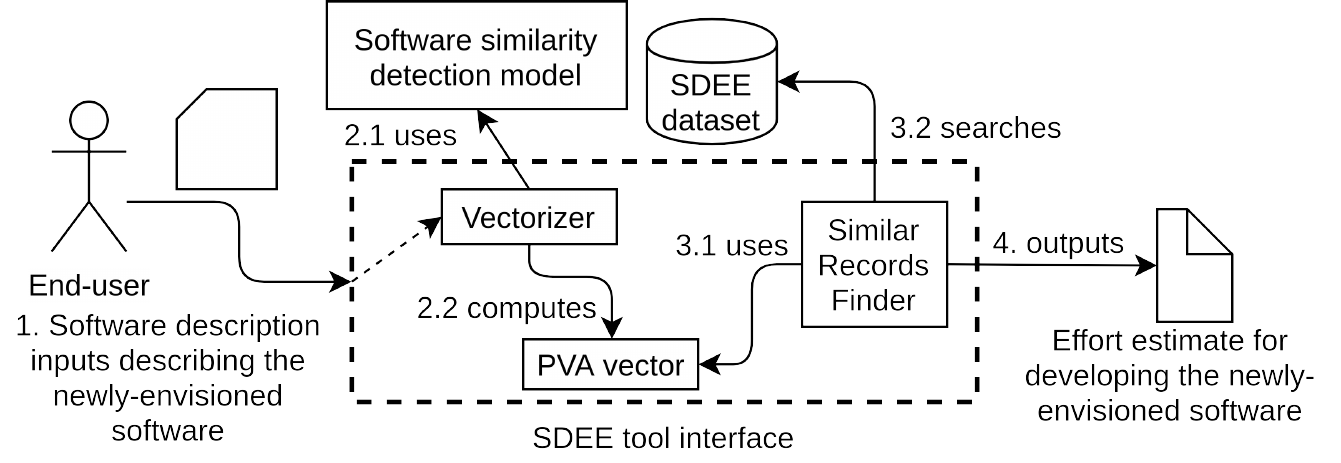}
\caption{Estimating the effort for developing a newly-envisioned software}
\label{fig:estimating-SDEE-using-tool-mindmap}
\end{figure*}
    \item \textbf{SDEE tool:} \label{sec:sdee-engine} Fig. \ref{fig:estimating-SDEE-using-tool-mindmap} highlights the major components of the tool. For a given newly-envisioned software $z$, our tool performs the following steps:
    \begin{enumerate}
        \item Using the \emph{vectorizer} module and the best-performing software similarity detection model $\hat{M}$, it converts the software project description inputs into a PVA vector $v$.
        \item Using $v$ as the input to the \emph{similar-records fetcher}, it fetches the top-k most similar software matches ($K$) to $v$ from the SDEE dataset. We use the cosine similarity\footnote{\url{http://bit.ly/2ODWoEy}} to measure the similarity between the PVA vectors.
        \item 
        To compute the effort estimate for developing $z$, it combines the effort estimates of $K$ by using the Walkerden's triangle function represented in equation \ref{eq:traingle-function}.
    \end{enumerate}
\end{enumerate}

\section{Performance Evaluation and Comparison}
\label{sec:performance-evaluation-and-comparison}
\begin{table}
    \centering
    \caption{Developer activity information used in developing the SDEE dataset}
    \label{tab:dataset-project-attributes}
    \resizebox{0.8\columnwidth}{!}
    {
    \begin{tabular}{c|c} 
    \toprule
     \textbf{Attribute name} & \textbf{Description}\\ 
     \midrule
     \multirow{2}{*}{\emph{\shortstack{Developer count\\ (devCount or $\mid D^{r} \mid$)}}} & \multirow{2}{*}{\shortstack{The total number of developers working on an OSS repository $r$. It is\\ measured in \emph{persons}.}}\\
     &\\
     \hline
    \multirow{2}{*}{\emph{$SLOC_m^{r}$}} & \multirow{2}{*}{\shortstack{The total number of SLOC modifications made in developing the OSS repository $r$.\\ It is measured in \emph{lines}, with the whitelines excluded.}}\\
    &\\
    \hline
    \multirow{2}{*}{\emph{\shortstack{Development time\\ $t^{r}$}}} & \multirow{2}{*}{\shortstack{The total amount of time spent in developing the OSS repository $r$.\\ It is measured in \emph{months}.}}\\
    &\\
    \hline
    \multirow{2}{*}{\emph{\shortstack{Development effort\\ $e^{r}$}}} & \multirow{2}{*}{\shortstack{The total amount of effort spent in developing the OSS repository $r$.\\ It is measured in \emph{person-months} and computed using equation \ref{eq:software-development-effort}.}}\\
     &\\
    \bottomrule
    \end{tabular}
   } 
\end{table}

A vital aspect of an SDEE method is detecting similar software matches to a newly-envisioned software whose effort is estimated.  Most of the existing SDEE methods, such as ATLM\cite{whigham2015baseline}, ABE \cite{kocaguneli2011exploiting},  LOC \enquote{straw man} \cite{menzies2017negative}, Artificial Neural Networks (NeuralNet) \cite{minku2013software}, use different methods to detect similar software matches using the metadata of past software projects available as existing datasets, such as PROMISE repository.
In contrast, we use the software project descriptions to develop our \emph{software similarity detection model} to perform this task. One of our key hypotheses is to validate that our \emph{software similarity detection model} trained using software product descriptions helps detect more relevant software matches than the models trained using project metadata. 

We perform several experiments using the  \emph{randomised trials method} (discussed in Section \ref{sec:exp-1}) and the \emph{k-fold cross-validation} method (discussed in Section \ref{sec:exp-2}) to validate our hypothesis by comparing the performance of our method with the above methods. We choose these methods for comparison as these are the most related works to ours and have been used as baselines in the last ten years by several prominent studies. 
The experiments' details to obtain an optimally tuned \emph{software similarity detection model} are provided in Appendix-\ref{ap:PVA-parameter-tuning}. Also, as ABE and LOC \enquote{straw man} use the k-nearest neighbors (kNN) method for detecting similar software matches, we compared the performance of these methods by experimenting with k=1, 2, 3, 4, and 5 for kNN in both the randomized trials and the cross-validation experiments. We used the scikit-learn python library \cite{scikit-learn} to implementing ATLM, NeuralNet, and kNN. 
To validate the significance of our results, we performed various statistical significance tests, such as t-test with bootstrapping and cliff's $\delta$ test \cite{cliff1993dominance,semenick1990tests} (discussed in Section \ref{sec:exp-3} and Section \ref{sec:exp-4}). To validate the correlation between the SDEE metrics (used to compute effort) and the software development effort, we test using Pearson's correlation coefficient \cite{benesty2009pearson} (discussed in Section \ref{sec:exp-5}).


Since both the randomized and cross-validation are examples of internal validation, and it has been considered essential to perform external validation before adopting the prediction models in practice \cite{bleeker2003external}, we perform Experiment \# 6 (discussed in Section \ref{sec:exp-6}). To perform these experiments, we invited professional programmers and asked them to evaluate our tool's behavior on various quality parameters, such as the response time, accuracy, and the ease-of-use of our tool.  The programmers were asked to use the existing software project estimates available as existing datasets such as COCOMO81\footnote{\url{https://bit.ly/3eMeFcN}} or by fetching the data from VCSs such as GitHub. For performing  this  experiment,  we  provided  access  to  our  tool by sharing it at \url{https://doi.org/10.5281/zenodo.5095723}. Table \ref{table:experiments-summary} summarizes the significant findings of our experiments.

\begin{table}
    \centering
    \caption{SDEE: Experiments summary}
    \label{table:experiments-summary}
    \resizebox{\columnwidth}{!}
    {
    \begin{tabular}{c|c|c} 
    \toprule
     \textbf{Experiment} & \textbf{Objectives} & \textbf{Major findings} \\ 
     \midrule
    
    \multirow{3}{*}{\shortstack{Experiment \#1\\ (Section \ref{sec:exp-1})}}& \multirow{3}{*}{\shortstack{Performance Comparison with\\ the existing SDEE methods using\\ the randomized training and test sets}}&\multirow{3}{*}{\shortstack{DevSDEE achieves the highest Standard Accuracy (SA)-\\ of 59.89\% over the random guessing method, 42.7\% with-\\ ATLM, and 84.22\% with the LOC straw man estimator.} }\\
    &&\\
     &&\\
    \hline
    \multirow{3}{*}{\shortstack{Experiment \#2\\ (Section \ref{sec:exp-2})}}& \multirow{3}{*}{\shortstack{Performance Comparison with\\ the existing SDEE methods using\\ the k-fold cross-validation method}}&\multirow{3}{*}{\shortstack{DevSDEE achieves the highest SA of 57.05\% over-\\ the random guessing method, 35.13\% with ATLM,\\ and 87.26\% with the LOC straw man estimator.} }\\
    &&\\
     &&\\
    \hline
    \multirow{3}{*}{\shortstack{Experiment \#3\\ (Section \ref{sec:exp-3})}}& \multirow{3}{*}{\shortstack{To validate the negligible difference\\ between the DevSDEE estimates\\ and the true effort values.}}&\multirow{3}{*}{\shortstack{t-significant test, Cliff's delta, and the parametric-\\ effect tests show the minimal differences\\ in the DevSDEE estimates and the true effort values.} }\\
    &&\\
     &&\\
    \hline
    \multirow{3}{*}{\shortstack{Experiment \#4\\ (Section \ref{sec:exp-4})}}& \multirow{3}{*}{\shortstack{To validate that DevSDEE provides\\ better effort estimates than the existing\\ methods using t-significance and effect tests}}& \multirow{3}{*}{\shortstack{Both t-tests and effect tests validate that\\ DevSDEE results in more accurate (less error)-\\ effort estimates than the existing methods.}} \\
     &&\\
    &&\\
    \hline
    \multirow{3}{*}{\shortstack{Experiment \#5\\ (Section \ref{sec:exp-5})}}& \multirow{3}{*}{\shortstack{Effect of SDEE metrics on the\\ overall development effort values}}& \multirow{3}{*}{\shortstack{Correlation value of +0.799 for developer- \\ count and +0.644 for development time\\  compared with the software effort values.}} \\
    &&\\
    &&\\
    \hline

    \multirow{3}{*}{\shortstack{Experiment \#6\\ (Section \ref{sec:exp-6})}}& \multirow{3}{*}{\shortstack{Performance evaluation from a programmer’s\\ perspective on various quality parameters,\\ such as ease of use, response time, and accuracy.}}& \multirow{3}{*}{\shortstack{SDEE achieves ratings $>8/10$\\ for most of the considered\\ quality parameters.}} \\
    &&\\
    &&\\
    \bottomrule
    \end{tabular}
   } 
\end{table}

\subsection{Test-bed Setup}


To perform the comparison, we extracted the developer activity information from $19,096$ releases of $1,587$ GitHub OSS repositories and created a dataset of projects' metadata attributes. We also extracted the respective project descriptions. Thus, the input source for both our method and the existing methods is the same, though the respective inputs are different, viz., the metadata information and the project descriptions.  After removing the records for repositories having time (in months) or developer-count $<$ 1, our dataset reduces to $1,184$ GitHub repositories' records. 
Table \ref{tab:dataset-project-attributes} provides the details of the project attributes used to develop the dataset. We want to emphasize that we use the same dataset to compute the effort values across different SDEE techniques. Therefore, the input remains the same for all the methods. 

\subsection{Evaluation metrics}
Most of the existing methods \cite{kocaguneli2011exploiting,minku2013software,menzies2017negative,whigham2015baseline}, with which we present our comparison, use the evaluation metrics such as the magnitude of the relative error (MRE), the mean magnitude of the relative error (MMRE), the median magnitude of the relative error (MdMRE), and prediction level (PRED) for performance evaluation \cite{conte1986software}. However, we also include some additional experimental evaluations based on newer software metrics, such as mean absolute error (MAR) \cite{langdon2016exact}, logarithmic standard deviation (LSD), RE*, standard accuracy (SA), and effect size \cite{shepperd2012evaluating}, as these are recommended by others \cite{kitchenham2002preliminary,miller2000applying}.

For the evaluation metrics defined below, let $i$ be a software project, $e$ be the actual effort value to develop $i$, and $e'$ be the estimated effort value obtained by an SDEE model $M$. 
Also, let $n$ be the total number of considered software projects.  
\begin{enumerate}
\item \emph{Magnitude of Relative Error, MRE:} It measures the relative size of the difference between the actual and estimated effort value:
  \begin{equation}
       \label{eq:mre}
    MRE =    \frac{(\mid e' - e \mid)}{e}
   \end{equation}
  The lower the MRE values for a model, the lesser is the error in its estimates, and hence, better is its performance.
  
  \item \emph{Mean MRE (MMRE):} : It measures the percentage of the MREs averaged over the n projects present in the test-bed:
  \begin{equation}
       \label{eq:mmre}
    MMRE =    \frac{100}{n} \sum_{i=1}^{n}{MRE}_i
   \end{equation}
  The lower the MMRE value for a model, the smaller is the averaged error in its estimates, and hence, better is its performance.
  
    \item \emph{Median MRE (MdMRE):} :: It measures the median of MRE values obtained by the model using the software projects in the test-bed:

  \begin{equation}
       \label{eq:mdmre}
    MdMRE =    Median(\forall_{i=1}^{n} {MRE}_i)
   \end{equation}
  The lower the MMRE value for a model, the smaller is the averaged error in its estimates, and hence, better is its performance.

\item \emph{Mean absolute residual, MAR:} It measures the mean of the absolute value of the residuals, i.e., the error between the predicted and the true value of an estimate, and is mathematically represented as:
  \begin{equation}
       \label{eq:mar}
    MAR =    \frac{\sum_{i=1}^{n}(\mid e'_i - e_i \mid)}{n}
   \end{equation}
  The lower the MAR values for a model, the lesser is the error in its estimates, and hence, better is its performance.
 \item \emph{Logarithmic standard deviation, LSD:} It measures standard deviation of the log transformed prediction values, and is mathematically represented as:
  \begin{equation}
       \label{eq:lsd}
   LSD = \sqrt{\sum_{i=1}^{n} (ln\ e'_i - ln\ \overline{e_i})^2}
   \end{equation}
   The lesser the LSD values, the lesser is the error, and the better is the performance.
   
\item \emph{RE*:} It is used as a baseline error measurement metric, and is mathematically represented as:
\begin{equation}
       \label{eq:re}
    RE* = \frac{var(residuals)}{var(measured)}   
   \end{equation}
where var represents the variance, residuals represents the error value, i.e., $e'-e$
, and the measured values represent the actual effort values, i.e., $e$ values. The lesser the $RE^*$ values for a model, the better is the performance.   
\item \emph{Standard Accuracy, SA:} It provides a relative assessment of the performance of the model by comparing it with the random guessing method, and is mathematically represented as:
\begin{equation}
       \label{eq:sa}
    SA_P = (1 - \frac{MAR_P}{\overline{MAR_{RG}}})*100   
   \end{equation}
where P represents the SDEE method whose performance is being evaluated, RG represents the random guessing method  \cite{shepperd2012evaluating}.
Over large number of runs, $MAR_{RG}$ converges to sample mean.
SA represents how much better an effort estimation technique is than random guessing. The larger the value of SA, the better is the model performance.

\item \emph{Effect size:} \label{pt:effect-size} It provides the  magnitude of the relationship between treatment variables and outcome variables and is computed based on the sample data to make inferences about a population (analogously to the concept of hypothesis testing) \cite{kampenes2007systematic}. It is further categorized into parametric and non-parametric based on the nature of the data. The parametric method assumes the data to have a normal distribution, whereas the non-parametric method is free from any such assumption. The parametric versions of effect size include cohen's $d$, hedges' $g$, and glass $\Delta$ \cite{cohen2013statistical,hedges2014statistical,hedges1981distribution}, while the non-parametric versions include cliff's $\delta$ effect size and bootstrap method \cite{cliff1993dominance}. The mathematical expressions of these variations are listed below:
\begin{equation}
       \label{eq:cohen's-effect-size}
    Cohen's\ d = (\overline{X_1} - \overline{X_2}).\sqrt{\frac{(n_1 -1).var_1 + (n_2 -1).var_2}{n_1 +n_2 -2}} = t.\sqrt{\frac{1}{n_1} + \frac{1}{n_2}}
   \end{equation}
where $n_1, n_2$ represent the sample size of the two groups in consideration ($n$ in the present case), $\overline{X_1}, \overline{X_2}$ represent the mean of the considered groups ($\overline{e'}$ and $\overline{e'}$ in the present case), and $var_1, var_2$ represent the variance of the values in the respective groups. The equivalent expression with t (from the t-test) was given by Rosenthal \cite{rosenthal1994parametric}. It was observed that for a small sample size ($<20$), cohen's effect size values are biased on the sample means of the population. To rectify this limitation, Hedges proposed a correction expressed as:
\begin{equation}
       \label{eq:hedges'-effect-size}
    Hedge's\ g = d.({1 - \frac{3}{4.(n_1 +n_2) - 9}}) 
   \end{equation}
where d is the cohen's effect size. Both d and g are considered as small effect sizes if the values are $<=0.2$, medium effect sizes if the values are $<=0.5$, and large effect sizes if values are  $>=0.8$.

Similarly, Glass's $\Delta$ is recommended for use when the standard deviations of the considered groups differ substantially:
\begin{equation}
       \label{eq:glass's-effect-size}
    Glass's\ \Delta = \frac{\overline{X_1} - \overline{X_2}}{s_{control}}
   \end{equation}
where $s_{control}$ represents the standard deviation of the control group, and

\begin{equation}
       \label{eq:cliff's-delta}
    Cliff's\ \delta = \frac{\#(X_1 > X_2) - \#(X_2 > X_1)}{n_1.n_2}
   \end{equation}
   where $\#$ represents the count or cardinality. For instance, $\#(X_1 > X_2)$ implies the count of data values in group 1, which are numerically larger (in value) than that of the corresponding data values in group 2. The $\delta$ value lies in the range [-1,1], with the effect size of -1 or 1 ,implying an absence of overlap of data values, whereas an effect size of 0 representing a complete overlap \cite{macbeth2011cliff}.
 
\end{enumerate}

\subsection{Experiment \#1: Randomized trials}
\label{sec:exp-1}
\textbf{Objective:} To evaluate the performance of SDEE models and compare them with the existing methods using the randomized trial method. 

\textbf{Procedure:} When evaluating models' performance using randomized trials with a dataset containing n rows (or feature vectors), x rows are selected without replacement for testing and the rest of the (n-x) rows for training. This process is repeated for r number of trials, with the rows being shuffled in each trial. x=10 is a suitable value  \cite{menzies2006selecting} and has been used in the literature  \cite{minku2013software, kocaguneli2011exploiting}. Some of the existing studies take r = 30 \cite{minku2013software}, whereas some validate their models using r = 20  \cite{kocaguneli2011exploiting}. With a dataset of 1184 rows, we selected x = 55 and r = 20 to have a considerable sample size for each iteration. Thus, 55 rows were chosen in each iteration without replacement for testing, and the remaining 1129 (=1184-55) rows for training.  

\begin{table}
\centering
\caption{Performance comparison based on the Randomized trials}
\label{tab:rand-trial-results}
 \resizebox{0.8\columnwidth}{!}
    {
\begin{tabular}{c|c|c|c|c|c|c|c}
\toprule

\multicolumn{2}{c|}{\multirow{2}{*}{\shortstack{SDEE\\ method}}} & \multicolumn{5}{c}{Evaluation metrics (mean $\pm$ std.dev)}\\
\cline{3-8}
\multicolumn{2}{c|}{}&$LSD$ & $RE^*$ & $MAR$ & $MMRE$ &$MdMRE$ & $SA$ \\ \midrule

\multicolumn{2}{c|}{ATLM}	&	2.08 $\pm$ 0.21	& \cellcolor{lightgray}	0.46 $\pm$ 0.33	&	 \cellcolor{lightgray} 34.82 $\pm$ 6.31	&	192.40 $\pm$ 55.69 & \cellcolor{lightgray} 40.07 $\pm$ 17.05  & \cellcolor{lightgray} 53.57 $\pm$ 8.54	\\ \hline

\multicolumn{2}{c|}{NeuralNet}	&	1.40 $\pm$ 0.62	&	4.31 $\pm$ 12.17	& \cellcolor{lightgray} 36.04 $\pm$ 26.96	& 222.78 $\pm$ 285.64	& \cellcolor{lightgray} 59.07 $\pm$ 16.65  & 9.51 $\pm$ 79.47\\ \hline

\multirow{3}{*}{ABE}&	neigh =1&	1.45 $\pm$ 0.63	&	1.69 $\pm$ 3.90	&	37.51 $\pm$ 47.21	&	557.74 $\pm$ 1968.9	&	557.74 $\pm$ 1968.9	&	9.69 $\pm$ 103.68	\\ \cline{2-8}
	&	neigh =2	&	1.08  $\pm$ 0.43	&	1.24  $\pm$ 2.57	&	37.19  $\pm$ 41.16	&	632.46  $\pm$ 1615.35	&	632.46  $\pm$ 1615.35	&	9.71  $\pm$ 90.20	\\ \cline{2-8}
	&	neigh =3	& \cellcolor{lightgray} 0.83 $\pm$ 0.46	&	\cellcolor{lightgray} 1.21 $\pm$ 2.82	&	\cellcolor{lightgray} 32.89 $\pm$ 42.28	&	596.31 $\pm$ 1429.21	&	596.31 $\pm$ 1429.21	& \cellcolor{lightgray}	20.41 $\pm$ 89.89	\\ \cline{2-8}
   &	neigh =4	& \cellcolor{lightgray}	0.83 $\pm$ 0.47	&	1.29 $\pm$ 3.09	&	35.20 $\pm$ 43.31	&	588.48 $\pm$ 1332.76	&	588.48 $\pm$ 1332.76	&	12.44 $\pm$ 90.94	\\ \cline{2-8}
	&	neigh =5	& \cellcolor{lightgray}	0.74 $\pm$ 0.41	&	1.40 $\pm$ 3.13	&	35.15 $\pm$ 43.01	&	582.21 $\pm$ 1218.80	&	582.21 $\pm$ 1218.80	&	14.27 $\pm$ 88.21	\\
	\hline
\multirow{3}{*}{\shortstack{LOC\\ straw\\ man}}	&	neigh =1	&1.38 $\pm$ 0.71	&	3.33 $\pm$ 7.36	&	76.37 $\pm$ 105.65	&	899.37 $\pm$ 1983.16	&	899.37 $\pm$ 1983.16	&	-90.39 $\pm$  278.33	\\
\cline{2-8}
	&	neigh =2	&	1.13 $\pm$ 0.47	&	5.50 $\pm$ 9.96	&	65.58 $\pm$ 75.53	&	941.74 $\pm$ 1628.10	&	941.74 $\pm$ 1628.10	&	-62.59 $\pm$ 179.23	\\
	\cline{2-8}
	&	neigh =3	&	0.93 $\pm$ 0.48	&	3.15 $\pm$ 5.98	&	58.50 $\pm$ 65.56	&	882.80 $\pm$ 1491.72	&	882.80 $\pm$ 1491.72	&	-44.7 $\pm$ 148.32	\\ 
	\cline{2-8}
	&	neigh =4	& \cellcolor{lightgray}	 0.82  $\pm$ 0.45	&	3.28  $\pm$ 5.98	&	54.28  $\pm$ 60.62	&	850.19  $\pm$ 1434.05	&	850.19  $\pm$ 1434.05	&	-34.4 $\pm$ 132.68	\\ 
	\cline{2-8}
	&	neigh =5	& \cellcolor{lightgray}	0.81 $\pm$  0.44	& \cellcolor{lightgray}	2.70 $\pm$  5.47	& \cellcolor{lightgray}	53.64 $\pm$  60.94	&	820.76 $\pm$  1387.48	&	820.76 $\pm$  1387.48	& \cellcolor{lightgray}	-32.83 $\pm$  133.53	\\ 
	\hline
 
 \multirow{3}{*}{DevSDEE}	&	 neigh =1	& 0.95 $\pm$ 0.19	&	0.52 $\pm$ 0.97	& \cellcolor{lightgray}	9.71 $\pm$ 8.69 & \cellcolor{lightgray}	130.61 $\pm$ 25.91	&	\cellcolor{lightgray} 119.77 $\pm$ 42.0		&	57.54 $\pm$ 10.52	\\	\cline{2-8}
&	neigh =2	& \cellcolor{lightgray} 0.14 $\pm$ 0.34 & \cellcolor{lightgray} 0.34 $\pm$ 0.41	& 9.82 $\pm$ 7.92 &	166.01 $\pm$ 40.78	&	153.40 $\pm$ 53.81	& \cellcolor{lightgray}	59.89 $\pm$ 6.88	\\ \cline{2-8}
&	neigh =3	& 0.18 $\pm$ 0.19	&	0.41 $\pm$ 0.28	&	16.01 $\pm$ 4.24 &	540.24 $\pm$ 238.88	&	524 .19 $\pm$ 264.18	&	47.79 $\pm$ 7.75	\\	\cline{2-8}
&	neigh =4	&	0.25 $\pm$ 0.10	&	0.47 $\pm$ 0.20	&	16.62  $\pm$ 3.65 &	662.20 $\pm$ 294.57	&	636.18 $\pm$ 335.09	&	39.97 $\pm$ 9.51	\\  \cline{2-8}
&	neigh =5	& 0.29 $\pm$ 0.06	&	0.57 $\pm$ 0.13	&	17.7 $\pm$ 2.98 &	768.11 $\pm$ 334.27	&	729.34 $\pm$ 393.88	&	33.76 $\pm$ 11.95	\\
\bottomrule

\end{tabular}
}
\end{table}

\textbf{Observations:} Table \ref{tab:rand-trial-results} shows the results obtained from these experiments. The averaged evaluation metrics values and their standard deviations (std. dev.) for 20 iterations are listed. The grey-colored cells in the table represent the SDEE models performing better than the rest.  As observed from the table values, DevSDEE outperforms the existing methods for the considered evaluation metrics. We obtain the best performing DevSDEE values at neigh = 2 with the highest SA of 59.89\% (error=$\pm$6.88).  However, the $SA$, $MAR$, $MMRE$, and $MdMRE$  values have large standard deviation (std. dev. or error) values and thus are not very reliable. Nevertheless, the $LSD$ and $RE^*$ values present a clear picture of DevSDEE performing better than the compared SDEE methods.   

\textbf{Critical Analysis:} As it is evident from the results, the $LSD$ and $RE^*$ values of ABE, and LOC Straw man estimator decrease (or improve) with the increase in the neighbor (neigh) count, whereas DevSDEE gives the best results for neigh = 2 after which the values increase. The difference in DevSDEE's behavior from the existing methods can be explained due to the difference in detecting similar software matches. To estimate the similar software matches for a newly-envisioned software $z$, DevSDEE detects the similarity in the software project descriptions of existing software and $z$. It is improbable to have multiple software with very similar software project descriptions. Hence, it is evident that only the effort values of few most-similar matches would be close to the effort value of $z$, and would contribute towards the SDEE process. However, for the existing methods working with project metadata values to detect similarity, it is possible to have a considerable number of software projects with similar project metadata attribute values. Thus, more are such similar software matches, better would be the final estimate value, and lesser would be the $LSD$ and $RE^*$ values. 

\textbf{Negative SA values:} The negative SA values for the LOC straw man estimator signify that the random guessing method performs better than it. However, since the values have large std. dev. (or error) values associated with them, we cannot reach any conclusion with confidence.

\subsection{Experiment \#2: k-fold cross-validation}
\label{sec:exp-2}
\textbf{Objective:} To evaluate the performance of SDEE models and compare with the existing methods using the k-fold cross-validation method.

\textbf{Procedure:} In k-fold cross-validation, the complete set of features is divided into k parts, with one part used for testing and the rest of the (k-1) parts for training. Thus, the complete training-testing phenomenon is performed for k number of times with different training and testing sets. Three-fold, five-fold, and ten-fold cross-validation experiments were performed to compare the considered SDEE methods' performance. Since the best results for most of the methods were obtained corresponding to the ten-fold cross-validation, we restrict our discussion to it. However, the complete set of results and models are shared at \url{https://bit.ly/3cuhRtQ}. 

\begin{table}
\centering
\caption{Performance comparison based on the Ten-Fold cross-validation}
\label{tab:k-fold-results}
 \resizebox{0.8\columnwidth}{!}
    {
\begin{tabular}{c|c|c|c|c|c|c|c}
\toprule

\multicolumn{2}{c|}{\multirow{2}{*}{\shortstack{SDEE\\ method}}} & \multicolumn{5}{c}{Evaluation metrics (mean $\pm$ std.dev)}\\
\cline{3-8}
\multicolumn{2}{c|}{}&$LSD$ & $RE^*$ & $MAR$ & $MMRE$ & $MdMRE$ & $SA$\\ \midrule

\multicolumn{2}{c|}{ATLM}	&	2.08 $\pm$ 0.21	& \cellcolor{lightgray}	0.48 $\pm$ 0.09	& \cellcolor{lightgray}	15 $\pm$ 2.82	&	192.4 $\pm$ 55.69	& \cellcolor{lightgray}	40.07 $\pm$ 17.05	& \cellcolor{lightgray}	53.57 $\pm$ 8.55	\\
\hline
\multicolumn{2}{c|}{NeuralNet}	& \cellcolor{lightgray}	1.8 $\pm$ 0.88	&	22.29 $\pm$ 65.26	&	61 $\pm$ 76.3	&	328.93 $\pm$ 461.79	&	75.56 $\pm$ 42.38	&	-33.99 $\pm$ 139.52	\\
\hline
\multirow{3}{*}{ABE}	&	neigh =1	&	1.1 $\pm$ 0.37	&	0.65 $\pm$ 0.86	&	28.96 $\pm$ 24.14	&	75.07 $\pm$ 68.04	&	75.07 $\pm$ 68.04	&	33.028 $\pm$ 53.78	\\ \cline{2-8}
	&	neigh =2	&	1.02 $\pm$ 0.49	& \cellcolor{lightgray}	0.51 $\pm$ 0.66	& \cellcolor{lightgray}	26.47 $\pm$ 22.56	& \cellcolor{lightgray}	51.52 $\pm$ 29.14	& \cellcolor{lightgray}	51.52 $\pm$ 29.14	& \cellcolor{lightgray}	38.73 $\pm$ 49.26	\\ \cline{2-8}
	&	neigh =3	&	0.99 $\pm$ 0.28	& \cellcolor{lightgray}	0.53 $\pm$ 0.79	&	\cellcolor{lightgray} 30.29 $\pm$ 24.83	&	76.01 $\pm$ 58.33	&	76.01 $\pm$ 58.33	&	29.93 $\pm$ 51.52	\\ \cline{2-8}
	&	neigh =4	&	0.91 $\pm$ 0.24	&	0.70 $\pm$ 0.98	&	33.20 $\pm$ 26.99	&	82.04 $\pm$ 72.82	&	82.04 $\pm$ 72.82	&	22.61 $\pm$ 58.31 	\\ \cline{2-8}
	&	neigh =5	& \cellcolor{lightgray}	0.74 $\pm$ 0.35	&	0.84 $\pm$ 1.23	&	33.98 $\pm$ 31.34	&	107.29 $\pm$ 120.88	&	107.29 $\pm$ 120.88	&	20.94 $\pm$ 70.27	\\ 
	\hline
\multirow{3}{*}{\shortstack{LOC\\ straw\\ man}}	&	neigh =1	&	0.97 $\pm$ 0.63	&	3.69 $\pm$ 7.93	&	64.21 $\pm$ 76.08	&	290.10 $\pm$ 505.84	&	290.10 $\pm$ 505.84	&	-54.8 $\pm$ 189.05	\\ \cline{2-8}
	&	neigh =2	&	0.9 $\pm$ 0.31	&	2.17 $\pm$ 3.63	&	53.98 $\pm$ 56.13	& \cellcolor{lightgray}	188.48 $\pm$ 237.89	& \cellcolor{lightgray}	188.48 $\pm$ 237.89	&	-28.66 $\pm$ 134.81	\\ \cline{2-8}
	&	neigh =3	& \cellcolor{lightgray}	0.71 $\pm$ 0.4	&	1.95 $\pm$ 3.25	&	50.77 $\pm$ 48.6	&	242.04 $\pm$ 418.77	&	242.04 $\pm$ 418.77	&	-21.9  $\pm$ 119.10	\\ \cline{2-8}
	&	neigh =4	& \cellcolor{lightgray}	0.67  $\pm$ 0.32	& \cellcolor{lightgray}	1.44  $\pm$ 2.0	&	47.44  $\pm$ 40.02	&	213.73  $\pm$ 357.44	&	213.73  $\pm$ 357.44	&	-12.99  $\pm$ 95.01	\\ \cline{2-8}
	&	neigh =5	& \cellcolor{lightgray}	0.6  $\pm$ 0.34	&	1.46  $\pm$ 1.94	& \cellcolor{lightgray}	46.16  $\pm$ 38.27	&	224.12  $\pm$ 366.30	&	224.12  $\pm$ 366.30	& \cellcolor{lightgray}	-9.74 $\pm$ 90.32	\\ 
\hline
\multirow{3}{*}{DevSDEE}	&neigh =1	&	0.69 $\pm$ 0.51	& \cellcolor{lightgray}	0.19 $\pm$ 0.01	& \cellcolor{lightgray}	9.47 $\pm$ 1.12	&	94.31 $\pm$ 25.29	& \cellcolor{lightgray}	54.58 $\pm$ 7.27 & \cellcolor{lightgray} 56.79 $\pm$ 2.41 \\ \cline{2-8}
	&	neigh =2	&	\cellcolor{lightgray}	0.29 $\pm$ 0.22	&	\cellcolor{lightgray} 0.41 $\pm$ 0.27	&	\cellcolor{lightgray} 8.46 $\pm$ 0.08	& \cellcolor{lightgray} 47.88 $\pm$ 23.2	&	\cellcolor{lightgray} 42.4 $\pm$ 23.2 & \cellcolor{lightgray} 57.05 $\pm$ 8.16 	\\ \cline{2-8}
	&	neigh =3	&	0.44 $\pm$  0.20	&	0.53 $\pm$  0.06	&	13.03 $\pm$ 0.67	&	\cellcolor{lightgray} 48.72 $\pm$ 27.45	& \cellcolor{lightgray}	48.72 $\pm$ 27.45	&	50 $\pm$ 16.92	\\ \cline{2-8}
	&	neigh =4	&	 0.43 $\pm$ 0.31		&	0.57 $\pm$ 0.80	&	17.22 $\pm$ 2.86	&		60.98 $\pm$ 15.24	& 60.98 $\pm$ 15.24	&	51.52 $\pm$ 21.79	\\ \cline{2-8}
	&		neigh =5	& \cellcolor{lightgray} 0.34 $\pm$ 0.21 &	0.70 $\pm$ 0.46	&	17.21 $\pm$ 0.44	&	49.17 $\pm$ 18.56	&	49.17 $\pm$ 18.56	&	50.37 $\pm$ 17.83	\\ 
\bottomrule

\end{tabular}}
\end{table}

\textbf{Observations:} Table \ref{tab:k-fold-results} shows the results obtained for ten-fold cross-validation experiments. DevSDEE achieves the lowest mean and standard deviation values for $LSD$, $RE^*$, $MAR$, $MMRE$, $MdMRE$, and $SA$ values and has the most negligible errors in its SDEE estimates. The grey-colored cells in the table represent the SDEE models performing attaining the best metrics' values.  As observed from the table values, DevSDEE outperforms the existing methods and achieves the highest $SA$ of 57.05$\pm$8.16 and least $RE^*$ of 0.19$\pm$0.01 with neigh=2 . 
Since the $MMRE$, $MdMRE$, and $SA$ values have large std. dev. values (or errors), we decide using the $MAR$, $LSD$ and $RE^*$ values of DevSDEE. The reason for the large std. dev. values could be their dependency on the central measures (mean or median) of the data values or the assumption of normal data distribution. Nevertheless, the $LSD$ and $RE^*$ values present a clear picture of DevSDEE performing better than the compared SDEE methods.  However, we perform the additional non-parametric statistical test, viz., t-test with the bootstrap procedure and the cliff's delta tests, that we discuss in Section \ref{sec:exp-3} and Section \ref{sec:exp-4}.

\textbf{Critical Analysis and Negative SA values:} The large std. dev. values in the case of $MMRE$, $MdMRE$, and $SA$ values signify the software effort estimates have a considerable variation. This could be due to the difference in the nature of software types and the kind of developer activity performed in developing them. The effort required to develop a specific software library might differ depending on their inherent functionalities or characteristics. For instance, the Google Draco\footnote{\url{https://github.com/google/draco}} library is used to compress and decompress 3D geometric meshes and point clouds, while the Google Tink\footnote{\url{https://github.com/google/tink}} library is used to perform encryption to provide secure APIs. Both the libraries differ in their functionalities and hence would differ in the effort required to develop them. However, the effort values could also differ for the same software library types. For instance, Google snappy\footnote{\url{https://github.com/google/snappy}} is another compression library, but it aims for very high speeds and reasonable compression instead of maximum compression. Hence, the effort required to develop Draco would differ from snappy as well but is expected to be closer to snappy than Tink. The actual effort values essentially depend on the amount of developer activity (in the form of $SLOC_m$) performed on a software project, and we only aim to achieve an estimate using this approach. The conclusions for critical analysis on DevSDEE behavior observed in Section \ref{sec:exp-1} holds here as well, as DevSDEE performs best with neigh=1 or neigh=2. Similarly, the inference for negative values of LOC straw man's SA values remains the same.

\subsection{Experiment \#3: Significance tests}
\label{sec:exp-3}
\textbf{Objective:} To validate that there exists a negligible difference between the actual effort values and the DevSDEE estimates

\textbf{Need for statistical tests with the bootstrap procedure:} Both the Randomized trials and k-fold cross-validation methods have been considered to be essential to assess the model prediction error and avoid misleading estimates due to dataset instability \cite{turhan2012dataset}. However, these methods have been remarked to produce a large variance for the datasets with outliers and are considered unsuitable for SDEE datasets. This might be a reason behind the large error values in the case of k-fold cross-validation results as compared to the randomized trials’ results (as shown in Table \ref{tab:rand-trial-results} and Table \ref{tab:k-fold-results}), as the number of iterations is larger in the case of randomized trials as compared to the k-fold cross-validation method.

An alternative solution to this problem as suggested by the existing studies \cite{whigham2015baseline}, is to test using many iterations to obtain a fair estimate of performance and allow a meaningful statistical comparison. This is particularly what is done by statistical tests with bootstrap procedures, and therefore, we perform additional experiments using t-test, cliff-delta test, and other parametric effect tests with bootstrapping to validate our findings.

\textbf{Description:} For DevSDEE to perform efficiently, it should yield effort estimates close to the true effort values. Therefore, we perform this experiment to validate that DevSDEE estimates are close to true effort values or negligible difference between them.  The objective of this experiment is very similar to the null hypothesis $H_0$ of the effect tests, which states that there lies no difference between the test group and control group data values. All these effect size variations (defined in Section \ref{sec:evaluation-metrics}) are studied with $H_0$ and a significance value ($p$).

\textbf{Procedure:} We used the researchpy\footnote{\url{https://researchpy.readthedocs.io/en/latest/ttest_documentation.html}} and the DABEST python libraries\footnote{\url{https://github.com/ACCLAB/DABEST-python}} to compute the values of t-test, cohen's $d$, hedges' $g$, glass's $\Delta$, and cliff's $\delta$ \cite{dabest}. Both the libraries compute the estimates with a 95\% confidence interval and by taking 5000 bootstrap samples. To compute the estimate values, we provide the DevSDEE effort estimates and the true effort values obtained from the Random Trials and the k-fold cross-validation experiments as discussed in Section \ref{sec:exp-1} and Section \ref{sec:exp-2}.  

\begin{table}
\centering
\caption{Significance Testing to evaluate the DevSDEE results}
\label{tab:significance-test}
 \resizebox{\columnwidth}{!}
    {
\begin{tabular}{c|c|c|c|c|c|c|c|c|c}
\toprule

\multicolumn{2}{c|}{\multirow{2}{*}{}} & \multicolumn{2}{c|}{t-significance test} & \multicolumn{2}{c|}{Cliff's delta effect test} & \multicolumn{4}{c}{Parametric effect tests}\\
\cline{3-10}
\multicolumn{2}{c|}{}&$t-val$ & $p-val$ & $cliff's\ \delta$ & $p-val$ & $cohen's\ d$ & $hedges'\ g$ &  $glass's\ \Delta$ & $p-val$\\ 
\midrule
\multirow{5}{*}{\rotatebox[origin=c]{90}{Random Trials}} & neigh=1 & 0.23 $\pm$ 0.29	&	0.77 $\pm$ 0.11	&	0 $\pm$ 0.17	&	0.63 $\pm$ 0.15	&	0.22 $\pm$ 0.22	&	0.13 $\pm$ 0.18	&	0.32 $\pm$ 0.25	&	0.67 $\pm$ 0.1	\\
\cline{2-10}
& neigh=2	&	0.18 $\pm$ 0.27	&	0.80 $\pm$ 0.11	&	0.02 $\pm$ 0.09	&	0.65 $\pm$ 0.06	&	0.18 $\pm$ 0.21	&	0.1 $\pm$ 0.17	&	0.23 $\pm$ 0.27	&	0.66 $\pm$ 0.07	\\
\cline{2-10}
& neigh=3	&	0.11 $\pm$ 0.35	&	0.83 $\pm$ 0.19	&	0.03 $\pm$ 0.16	&	0.6 $\pm$ 0.11	&	0.10 $\pm$ 0.25	&	0.06 $\pm$ 0.22	&	0.12 $\pm$ 0.36	&	0.62 $\pm$ 0.11	\\
\cline{2-10}
& neigh-4	&	0.14 $\pm$ 0.4	&	0.81 $\pm$ 0.21	&	0.03 $\pm$ 0.16	&	0.60 $\pm$ 0.11	&	0.12 $\pm$ 0.28	&	0.08 $\pm$ 0.25	&	0.17 $\pm$ 0.43	&	0.6 $\pm$ 0.11	\\
\cline{2-10}
& neigh=5	&	0.15 $\pm$ 0.38	&	0.8 $\pm$ 0.20	&	0.03 $\pm$ 0.16	&	0.6 $\pm$ 0.11	&	0.13 $\pm$ 0.27	&	0.09 $\pm$ 0.23	&	0.25 $\pm$ 0.48	&	0.61 $\pm$ 0.09	\\
\hline
\multirow{5}{*}{\rotatebox[origin=c]{90}{k-Fold}} & neigh=1	&	0.06 $\pm$ 0.61	&	0.7 $\pm$ 0.24	&	0 $\pm$ 0.37	&	0.52 $\pm$ 0.31	&	0.05 $\pm$ 0.43	&	0.04 $\pm$ 0.37	&	0.07 $\pm$ 0.43	&	0.68 $\pm$ 0.23	\\
\cline{2-10}
& neigh=2	&	0.04 $\pm$ 0.58	&	0.70 $\pm$ 0.2	&	0.08 $\pm$ 0.19	&	0.62 $\pm$ 0.12	&	0.03 $\pm$ 0.41	&	0.02 $\pm$ 0.35	&	0.02 $\pm$ 0.48	&	0.67 $\pm$ 0.16	\\	\cline{2-10}
& neigh=3	&	0.13 $\pm$ 0.79	&	0.56 $\pm$ 0.16	&	0.12 $\pm$ 0.33	&	0.44 $\pm$ 0.08	&	0.09 $\pm$ 0.56	&	0.08 $\pm$ 0.49	&	0.02 $\pm$ 0.79	&	0.51 $\pm$ 0.18	\\	\cline{2-10}
& neigh=4	&	0.24 $\pm$ 0.88	&	0.48 $\pm$ 0.05	&	0.12 $\pm$ 0.33	&	0.44 $\pm$ 0.08	&	0.17 $\pm$ 0.62	&	0.15 $\pm$ 0.54	&	0.07 $\pm$ 0.96	&	0.43 $\pm$ 0.06	\\	\cline{2-10}
& neigh-5	&	0.24 $\pm$ 0.85	&	0.5 $\pm$ 0.06	&	0.12 $\pm$ 0.33	&	0.44 $\pm$ 0.08	&	0.17 $\pm$ 0.6	&	0.15 $\pm$ 0.52	&	0.09 $\pm$ 1.04	&	0.47 $\pm$ 0.06	\\	
\bottomrule

\end{tabular}
}
\end{table}

\textbf{Observations and Analysis:} Both the t-test and the parametric effect tests measure the difference between the mean values of these groups, and the $p$ value represents the significance (or probability) of these estimates. A large effort value associated with a $p<0.05$ (marked as 95\% confidence level) leads to the rejection of $H_0$, concluding that there is a considerable difference between the values of the considered groups. Similarly, if the effort values are small and are associated with large $p$ values, it is concluded that there is considerable overlap in the data values of the considered group, or the differences between the values are small \cite{cohen1988statistical}. However, $H_0$ can either be rejected or stated that it is impossible to reject the null hypothesis (if $p$ values are large or $>0.1$). As observed from Table \ref{tab:significance-test} values, all the t-test values have t values $<0.24 \pm 0.88$ associated with high $p$ values lie in $[0.48,0.83]$. A similar trend is observed for all the parametric effect tests, viz.,  $cohen's\ d$, $hedges'\ g$, and $glass's\ \Delta$, where all the effect values in the range $[0.02, 0.32]$ and the p-values in $[0.43, 0.68]$, which imply a small effect (as discussed in Section \ref{sec:evaluation-metrics} point \ref{pt:effect-size}). Also, the $cliff's \delta$ values range lies in $[0 \pm 0.17, 0.12 \pm 0.33]$ with large p values $[0.44, 0.65]$, which again depicts a large overlap between the DevSDEE estimates and true effort values. Therefore,  with such small t-values and effect values and large p-values, we can conclude that the differences between the estimates are minimal, and thus, we fail to reject the null hypothesis $H_0$. However, to validate our objective with better confidence, we perform additional experiments as discussed in Section \ref{sec:exp-4}.

\subsection{Experiment \#4}
\label{sec:exp-4}

\textbf{Objective:} To validate that DevSDEE provides better effort estimates than the existing methods

\textbf{Description:} To validate our hypothesis, we consider the $MAR$ of the estimates as inputs to the t-test and the effect experiments. While comparing DevSDEE with an existing approach $Z$, a large positive difference (effect or t values) between the $MAR$ values of the $Z$ estimates and the DevSDEE estimates would imply an increased error in $Z$ estimates as compared to DevSDEE's estimates. Similarly, we use the $SA$ measure for comparing the two estimates where we replace the $MAR$ of random guessing approach with the $MAR$ values of $Z$ in equation \ref{eq:sa}. The null hypothesis $H_0$ for this experiment would be defined as \emph{there lies negligible difference between the error ($MAR$) values of the estimates made by $Z$ approach and DevSDEE}, with $Z$ = \{ATLM, NeuralNet, ABE, LOC straw man\}.   

\textbf{Procedure:} We perform the t-test, cliff's $\delta$ test, and the parametric effect tests, viz., cohen's $d$, hedges' $g$, glass's $\Delta$ using the $MAR$ values obtained in the experiments \#1 and \#1 ( discussed in Section \ref{sec:exp-1} and Section \ref{sec:exp-2}). Additionally, we also compute the $SA$ measure values using the $MAR$ values as input. We use the researchpy\footnote{\url{https://researchpy.readthedocs.io/en/latest/ttest_documentation.html}} and the DABEST python libraries\footnote{\url{https://github.com/ACCLAB/DABEST-python}} to perform these tests. All the tests are performed with a 95\% confidence interval and by taking 5000 bootstrap samples.

\begin{table}
\centering
\caption{Significance Testing for comparison with the existing methods}
\label{tab:comparison}
 \resizebox{\columnwidth}{!}
    {
\begin{tabular}{c|c|c|c|c|c|c|c|c|c|c}
\toprule

\multicolumn{2}{c|}{\multirow{2}{*}{}} & \multirow{2}{*}{SA} &  \multicolumn{2}{c|}{t-significance test} & \multicolumn{2}{c|}{Cliff's delta effect test} & \multicolumn{4}{c}{Parametric effect tests}\\
\cline{4-11}
\multicolumn{2}{c|}{Method}&&$t-val$ & $p-val$ & $cliff's\ \delta$ & $p-val$ & $cohen's\ d$ & $hedges'\ g$ &  $glass's\ \Delta$ & $p-val$\\  
\midrule
\multirow{4}{*}{\rotatebox[origin=c]{90}{\shortstack{Random\\ Trials}}} & ATLM	&	42.7	&	2.49	&	0.02	&	0.74	&	$<0.00001$	&	1.06	&	1.02	&	1.74	&	0.02	\\
\cline{2-11}
& NeuralNet	&	73	&	4.13	&	0.0002	&	0.86	&	$<0.00001$	&	1.31	&	1.28	&	0.98	&	$<0.00001$	\\
\cline{2-11}
& ABE	&	72.66	&	2.58	&	0.01	&	0.57	&	0.0006	&	0.82	&	0.8	&	0.59	&	0.001	\\
\cline{2-11}
& LOC straw man	&	84.22	&	2.81	&	0.008	&	0.88	&	$<0.00001$	&	0.89	&	0.87	&	0.63	&	$<0.00001$	\\
\hline
\multirow{4}{*}{\rotatebox[origin=c]{90}{k-Fold}} & ATLM	&	35.13	&	2.46	&	0.02	&	0.74	&	$<0.00001$	&	0.78	&	0.76	&	1.87	&	0.01	\\
\cline{2-11}
& NeuralNet	&	56.35	&	1.38	&	0.05	&	0.52	&	0.04	&	0.62	&	0.59	&	0.45	&	0.05	\\	\cline{2-11}
& ABE	&	74.06	&	2.58	&	0.01	&	0.57	&	0.0006	&	0.82	&	0.8	&	0.59	&	0.001	\\	\cline{2-11}
& LOC straw man	&	87.26	&	2.81	&	0.008	&	0.88	&	$<0.00001$	&	0.89	&	0.87	&	0.63	&	$<0.00001$	\\	
\bottomrule

\end{tabular}
}
\end{table}

\textbf{Observations and Analysis:} The results of this experiment are listed in  Table \ref{tab:comparison}. As observed from the t-significance test results, the t-values lie in $[1.38, 2.81]$ with p-values $<=0.05$. Therefore, we can state that we can reject $H_0$ with a 95\% confidence level (and more in some cases where $p<0.05$). Hence, this also implies that DevSDEE estimates are considerably closer to the true effort values than the existing methods ($Z$).  
A similar trend is observed for parametric effect tests with effect values ranging in $[0.45, 1.74]$ and p-values $<=0.05$, which represents the values ranging from medium to a  effect (as stated in Section \ref{sec:evaluation-metrics} point \ref{pt:effect-size}). Thus, the parametric tests also reject $H_0$ with a 95\% confidence level. When analyzing the cliff's $\delta$ effect test results, the effect values range in $[0.52,0.88]$ with p-values $<=0.04$ (mostly $<0.001$). Thus, these effect values provide a 96\% level of confidence to reject $H_0$.

\subsection{Experiment \#5}
\label{sec:exp-5}

\textbf{Objective:} To validate the correlation between Software Development Effort Estimation metrics and software development effort

\textbf{Description:} We performed various experiments to validate our hypothesis that: 
 \textit{The developer activity information is useful in estimating the software development effort,} 
or more specifically, 
to validate that
\textit{SDEE metrics capture the developer activity information and show correlation with the overall development effort.} 


\textbf{Procedure:} To perform these experiments, we compute the correlation between various SDEE metrics values present in our dataset. We use the Pearson's correlation coefficient  \cite{benesty2009pearson}, which is used to measure the linear relationship between two variables. Note that Pearson's correlation coefficient has a range of values as [-1, 1].

To detect the correlation between different SDEE metrics, we computed Pearson's correlation coefficient values for the following SDEE metrics:
\begin{enumerate}
    \item[A)] The total number of SLOC modifications ($SLOC_m$) performed while developing a software repository vs.  the effort expended in developing it.
    \item[B)] The total number of developers (devCount) working on a software repository vs.  the effort expended in developing it.
    \item[C)] The total time spent developing a software repository vs.  the effort expended in developing it.
    \end{enumerate}

\begin{figure*}
\centering
\includegraphics[width=.5\columnwidth]{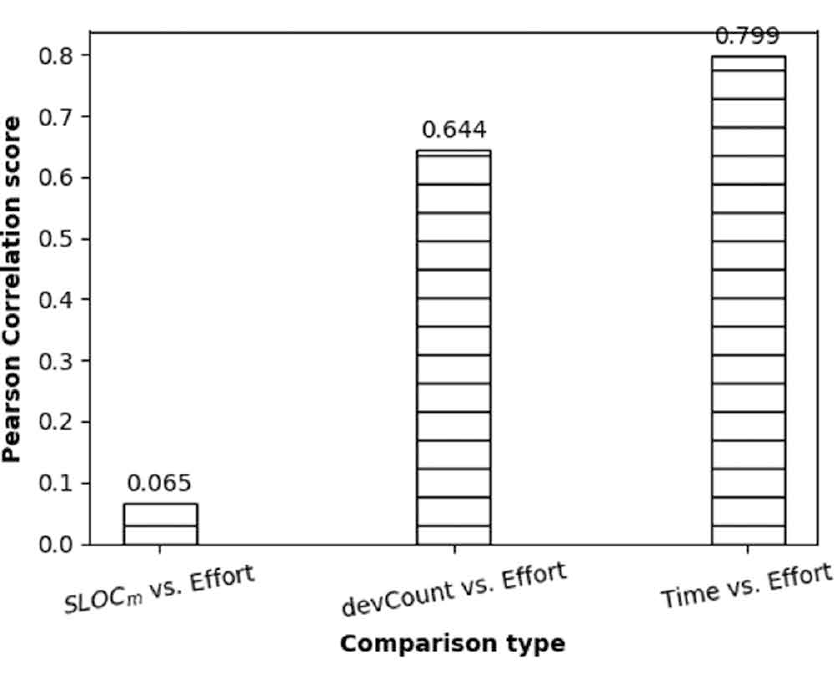}

\caption{Correlation between various SDEE metrics}
\label{fig:correlation}
\end{figure*}

\textbf{Results and observations:} Fig. \ref{fig:correlation} presents the correlation values for each of the comparison types listed above. As observed from the figure, development time has the highest correlation value of 0.799, which is explainable as more is the time spent developing a software project, more will be the effort (in person-months). Similarly, devCount shows a correlation value of 0.644 compared with the development effort and is explainable using the same reason for development time. However, $SLOC_m$ shows a small positive correlation value of $0.065$, which signifies that through the effort increases with the increase in $SLOC$ modifications, the increase is much smaller than the increase in devCount and development time. Thus, we can infer that devCount and development time are useful SDEE metrics and influence the software development effort more than the $SLOC_m$, which is the commonly used SDEE metric.   
\subsection{Experiment \#6}
\label{sec:exp-6}

\textbf{Objective:} To validate the performance of our SDEE tool by professional programmers and test if it works as intended to work for the user audience.

\textbf{The research question addressed:}  \emph{How does the SDEE tool perform from a programmer's perspective when judged on various quality parameters?}

\textbf{Description:} To evaluate our tool's performance, we consider the following quality parameters: the \emph{ease-of-use}, \emph{response-time}, and \emph{accuracy}. We invited professional programmers to perform this experiment and asked them to use our tool and validate its performance using the available information of existing GitHub repositories.  One hundred and eleven programmers responded. We had a mixture of candidates from final year computer science undergraduates (53), postgraduates (47), and IT industry professionals (11) with experience between 0 and 6 years. Each of the participants had prior knowledge of the software engineering fundamentals. The experiment was performed in  a controlled industrial environment.

\textbf{Procedure:} 
\label{sec:exp-4-procedure}
The programmers were requested to perform the following steps:

\begin{enumerate}
    \item Enter the details of a known or an available software product using the tool's input interface. The participants could use the available information of existing GitHub repositories to perform this validation and are not required to compute the effort values themselves.   
    \item Check the estimates provided as the response by the tool for the entered software requirements description by comparing it with the original effort values of the product. Note: We do not expect the participants to compute the effort values on their own. The participants were asked to use the effort values available online and validate that given a similar software description if the tool outputs a similar effort value. For instance, the participants could use the effort information of the existing GitHub repositories.
    \item Rate the tool's ease of use, response time, and accuracy, as observed from the experiment(s). The programmers were asked to provide their ratings on a scale of $1$--$10$. Here, $1$ represents the worst performance, and $10$ represents the best. 
\end{enumerate}

    \textbf{Interpretation of Table \ref{tab:experiment6-ratings}:} Table \ref{tab:experiment6-ratings} gives the details of the ratings received by the tool on various quality parameters as evaluated by the participants.  The scores represented in the table are on a scale of $1$--$10$, where $1$ represents the worst performance, and $10$ represents the best performance. Similarly, the votes are represented in the form of percentages. The sum of votes across every column equals $100$ (or \%). The first row of the table shows that a score of $10$ was awarded for \textit{ease-of-use} by $60\%$ voters, for \textit{response time} by $80\%$ voters, and for \textit{accuracy}, it was awarded by $60\%$ voters. Similarly, the third row says that the score of $8/10$ was awarded for all the quality parameters by only $10\%$ voters. 

\begin{table*}[ht]
\caption{Ratings recorded in Experiment \#6}
\label{tab:experiment6-ratings}
\centering
\resizebox{0.4\columnwidth}{!}{
\begin{tabular}{c|c|c|c} 
 \toprule[1pt]
 \textbf{Score $Y$}&\multicolumn{3}{c}{\textbf{Votes for the score $Y$ (in \%)}}\\
 \cline{2-4}
 (scale 1-10) &\emph{ease-of-use}&\emph{response time}&\emph{accuracy}\\
 \midrule[1pt]
 $10$ & $60$ & $80$ & $60$\\
 $9$ & $30$ & $10$ & $30$\\
 $8$ & $10$ & $10$ & $10$\\
 $7$ & $0$ & $0$ & $0$\\
 $6$ & $0$ & $0$ & $0$\\
 \bottomrule[1pt]
\end{tabular}
}
\end{table*}

\textbf{Results and observations:}
\label{sec:exp-6-results}
Some of the significant observations from Table \ref{tab:experiment6-ratings} are:

\begin{enumerate}
    \item Our tool received a rating of $>=8/10$ from all the participants for ease of use. Thus, most participants found it easy to use (or learn how to use it). 
    \item Our tool achieves impressive (or minimal) response time as it received a rating of $10/10$ by 80\% of the participants.
    \item All the participants found our tool's estimates to be considered accurate as they rated it $>=8/10$ for accuracy.
\end{enumerate}

\section{Threats to validity}
\label{sec:threats-to-validity}

\subsection{Threats to internal validity}
\label{sec:internal-validity}

\begin{enumerate}
    \item \emph{Handling the Corner cases:} A \enquote{refactoring} stage may occur in a software's life-cycle, producing a sudden spike in the modification and commits count. We treat such cases as the \enquote{corner} cases for our work. 
    
    \item \emph{Removing the outliers:} While developing the SDEE dataset, we performed the following to avoid the outliers:
    \begin{enumerate}
        \item Selecting OSS repositories having size $>=5$ MBs: We observed that the repositories with size $<5$ MBs were either at the very initial stage of software development or had very few source code files present in them. 
        \item Selecting the repositories that have been updated at least once in the last three years: To avoid the very old or inactive repositories. 
        \item Selected the repositories with more than $500$ stars: To fetch repositories with considerable developer involvement. 
        \item Filtering the repositories belonging to specific categories: To ensure the homogeneity of the dataset, we selected software repositories belonging to different types.
    \end{enumerate}
     \item \emph{Missed repositories:} Due to this constrained repository selection, we might have missed some relevant unlabeled (by categories) repositories or some relevant repositories with smaller star counts. However our objective of developing a homogeneous dataset (for software types) free from outliers is fulfilled. Nevertheless, as is the case with any experiment, it is impossible to cover all possibilities for the detection and avoidance of the swamping effect \cite{chiang2007masking}.
     
     \item \emph{Validity of the OSS project descriptions:} For developing our software effort estimation method, we determine similar software matches using a PVA model trained on the software product descriptions. Thus, we assume that the project descriptions present in VCS repositories are valid and do not change very often. The OSS software products considered while developing our tool are present in the GitHub OSS repositories and usually serve as software modules in developing more extensive software systems. Therefore, our system can only provide effort estimates to develop such modules of the software. The effort estimates provided corresponding to such modules of software modules can be summed up to obtain the aggregate effort estimate required to develop the more sophisticated software.

\end{enumerate}



\subsection{Threats to external validity}
\label{sec:external-validity}
Our work is based on the developer activity information extracted from various GitHub OSS repositories. Therefore, our experiments' results may differ when tested on proprietary software because industries have full-time working employees \cite{capiluppi2013effort}. Thus the developer activity while developing a proprietary software might differ due to the different working patterns. However, our approach remains valid regardless of whether we used the data from OSS or non-OSS projects.  For example, if the data used is taken from non-volunteer-driven projects such as proprietary software development, the approach would remain valid. This is because the similarity in software projects is detected based on the software descriptions and not the developer activity.

\subsection{Threats to statistical conclusion validity}
\label{sec:statistical-validity}
To validate the efficacy of our approach over the existing methods, we perform several statistical experiments as discussed in Section \ref{sec:exp-3} and Section \ref{sec:exp-4}. We report our findings at a 95\% confidence level with a 0.05 significance level. However, for many cases, we also obtained the results with a 99.999\% confidence level ($<0.00001$ significance).  


\section{Conclusion and Future Directions}
\label{sec:conclusion}
 SDEE for a software project is performed by analyzing the information of past software projects of similar nature. We propose a tool that estimates the effort required to develop a newly-envisioned software project $z$ by analyzing the developer activity information (tracked in VCSs) of software projects with similar functionality as $z$. SDEE metrics that we proposed are used to extract the developer activity information present in various VCS repositories. To determine the functionally similar software projects to $z$, we utilize the project descriptions associated with them. 

Our experimental results show that the developer activity metrics serve as a better input to an SDEE method than the existing software metrics, such as SLOC. Further, when trained on software product descriptions, a PVA-based document comparison allows for discovering more relevant software projects from the past whose effort estimates allow for more accurate and reliable SDEE. 
For instance, when estimating effort for a compression library ($z$) with \enquote{high compression} and \enquote{efficiency} as characteristics (used as text keywords by PVA for similarity detection), the top-2 similar matches obtained are Google Draco and Centaurean Density libraries. Both these libraries have the same characteristics highlighted in their project descriptions, and match the software requirements of $z$.
Our system achieves the highest standardized accuracy of 87.26\% (cliff's $\delta$=0.88)  when compared with the LOC straw man estimator and 42.7\% (cliff's $\delta$=0.74) with the ATLM method at 99\% confidence level ($p<0.00001$). Based on our experimental findings, we conclude that the use of developer activity information of past software helps develop better software estimation methods and tools. We plan to extend our work for providing incremental effort estimates (after one month, two months, and so on) in the future.
Since developer acts as the primary source of source-code contributor, we plan to study novel developer-related factors, such as developer's characteristics \cite{stray2017daily,yang2020developer,li2012leadership}, developer's geographical location \cite{rastogi2018relationship}, social interactions of developers \cite{iden2018social,zhang2017social}, and test their effect on software development.

\bibliographystyle{plainnat}
\bibliography{bibliography}

%% file: appendix.tex
\appendix
\section{Appendix}

\subsection{PVA parameter tuning experiments}
\label{ap:PVA-parameter-tuning}
The performance of PVA depends on its input parameters, viz., $\beta, \gamma,$ and $\psi$ (see Table \ref{tab:TableOfNotation}). 
Further, one of the significant challenges is determining the threshold similarity value ($\hat{\alpha}$) between two PVA vectors, which helps us classify them as similar or dissimilar.
We performed several experiments to determine the optimal values of $\beta, \gamma, \psi$, and $\hat{\alpha}$. Table \ref{table:training-scenarios} lists the experimental scenarios, while Algorithm-\ref{alg:training-pv-model} describes the specific steps performed for training the software similarity detection models. 

\begin{table}
    \centering
    \caption{Scenarios for training the software similarity detection models}
    \label{table:training-scenarios}
    \resizebox{0.8\columnwidth}{!}
    {
    \begin{tabular}{|c|c|c|c|} 
    \cline{1-3}
    \multicolumn{3}{|c|}{\textbf{Parameters varied}} & \\ \hline
    \textbf{Epochs, $\beta$} & \textbf{Vector size, $\gamma$}    & \textbf{Training samples, $\psi$} & \multicolumn{1}{c|} {\textbf{Models}}\\ \hline
    Fixed at 10 & Fixed at 10 & \multirow{2}{*}{\shortstack{Vary 0 to \texttt{CorpusSize}\\ in steps of 30}}& \multicolumn{1}{c|}{\texttt{CorpusSize} $\div$ 30} \\
    &&&\\
    \hline
    Vary 5 to 50 in steps of 5 & Fixed at 10 & Fixed at \texttt{CorpusSize} & \multicolumn{1}{c|}{10}\\ \hline
    Fixed at 10 & Vary 5 to 50 in steps of 5 & Fixed at \texttt{CorpusSize} & \multicolumn{1}{c|}{10} \\ \hline
    \end{tabular}}
\end{table}

\begin{algorithm}
\caption{Procedure to train PVA models}
  \label{alg:training-pv-model}
\begin{algorithmic}[1]
\REQUIRE  
$ H := $ Collection of software project descriptions.\\ 
$P := $ Set  of PVA parameter variation scenarios shown in Table \ref{table:training-scenarios}.\\

\ENSURE $M := $ The set of software similarity detection models trained using $H$ and $P$.\\

\FORALL{$\pi  \in P$}
    \STATE {$M(\pi):= trainPVAModel(H, \pi)$} \COMMENT{Uses gensim OSS library that implements PVA}
    \STATE {Save $M(\pi)$ to disk}
\ENDFOR
\end{algorithmic}
\end{algorithm}

\subsubsection{Test-bed set up}
\label{sec:test-bed set up}
To perform these experiments, we set up a test-bed $Y$ using the software product descriptions collected from GitHub repositories. To perform the validation testing of our software description similarity detection model, we partition the software project descriptions data into $\langle train\ data, test\ data \rangle$ in ratio 2:1. 
$Y$ comprises of $\langle same,different \rangle$ software product description pairs in equal proportion, viz., 50:50. 

\subsubsection{Evaluation metrics}
\label{sec:evaluation-metrics}

To evaluate the performance of our software description similarity detection model, we compute the F1 score\footnote{\url{https://bit.ly/3mZ70Ns}} and ROC area\footnote{\url{https://bit.ly/2Ev8BZP}} metrics defined as follows:

\begin{enumerate}
     \item \textbf{F1\ Score:} It is defined as the harmonic mean of precision and recall. 
       \begin{align}
       \label{ed:F1-score}
    F1\ score = \frac{2\times Precision \times Recall}{Precision + Recall}\end{align}
    
    where
    \begin{align}
    \label{eq:precision}
    Precision = \frac{TP}{TP + FP}
    \end{align}
    \begin{align}Recall = \frac{TP}{TP + FN}\end{align} TP=True positive, FP = False positive, and FN = False negative.
    \item \textbf{ROC area under the curve (AUC):} It measures the quality of the output. ROC curve is a plot that features the true positive rate (marked as Y-axis) vs. the false positive rate (marked as X-axis) for an experiment. 
\end{enumerate}

 Since the highest \emph{ROC curve area} value and the \emph{F1 score} value may differ across the models, we take the average of the two as the final accuracy measure of a model.


\subsubsection{Objective:} The prime objective of this experiment is to address the research question (RQ):

\emph{What is the effect of PVA parameters in performing the software similarity detection task?}

Some of the sub-RQs addressed under this are:
\begin{enumerate}[label=(\Alph*)]
    \item \emph{What is the highest accuracy achieved when detecting software similarity using our models?}
    \item \emph{What is the threshold similarity score between two PVA vectors for reliably detecting two software products as similar?}
    \item \emph{What is the effect of the parameters $\beta, \gamma, \psi$ on the performance of the software description similarity detection model?}
    \item \emph{What are the optimal values of $\beta, \gamma, \psi$ for training the software description similarity detection model?} 
\end{enumerate}

\begin{algorithm}[ht]
\caption{Procedure to test the software description similarity detection models trained using PVA}
  \label{alg:procedure-to-test-models}
\begin{algorithmic}[1]
\REQUIRE 

$ Y := $ Experimental test dataset containing $\langle same, different \rangle$ software product description pairs developed using the test-partition of the project descriptions.\\ 
$M := $ The set of software description similarity models trained using Algorithm \ref{alg:training-pv-model}.

\ENSURE $S_{same},  S_{different}:=$ Similarity score collections for $\langle same, different \rangle$ software product description pairs in $Y$.

\STATE{$S_{same} = S_{different} = \phi$}

\FORALL{project description pairs $\langle p_i, p_j \rangle \in Y$ }
    \FORALL{$\pi  \in P$}
        \STATE {$\phi_{i} = inferVector(p_i, M(\pi)) $} \COMMENT{Infer vector vectors using $M(\pi)$}
        \STATE {$\phi_{j} = inferVector(p_j, M(\pi)) $}

        \STATE {$\alpha = computeCosineSimilarity({\phi}_i, {\phi}_j) $}

        \IF{$p_i == p_j$}
            \STATE {$S_{same} = S_{same} \cup \langle \alpha \rangle$}
        \ELSE
            \STATE {$S_{different} = S_{different} \cup \langle \alpha \rangle$}
        \ENDIF
    \ENDFOR
\ENDFOR
 \end{algorithmic}
\end{algorithm}

\subsubsection{Procedure}
\label{sec:proc-exp-1}
\begin{enumerate}
    \item Obtain the set of similarity scores ($S$) for the collection of test pairs present in $Y$  using Algorithm \ref{alg:procedure-to-test-models}.
    
    \item \label{step:record-threshold-sim-scores} The $avg(S)$ value is stored as the threshold similarity score value ($\hat{\alpha}$). Note: the $\hat{\alpha}$ is computed separately for the \emph{same} and \emph{different} software product description pairs. Table \ref{table:thresholds} lists the threshold similarity score for the \emph{same} software product description pairs. 
    \item For each parameter combination $\pi \in P$:
    \begin{enumerate}
        \item Obtain the software description similarity detection model $\hat{M^{\pi}}$ such that, $\hat{M^{\pi}}$ achieves the highest evaluation scores among all $M^{\pi} \in M$. 
        \item Record the values of Accuracy scores and F1 scores obtained with $\hat{M^{\pi}}$ when tested on $Y$, and the tuning parameter combination ($\pi$) used in developing $\hat{M^{\pi}}$.
\end{enumerate}
\end{enumerate}

\begin{figure*}
\centering
\includegraphics[width=0.5\columnwidth]{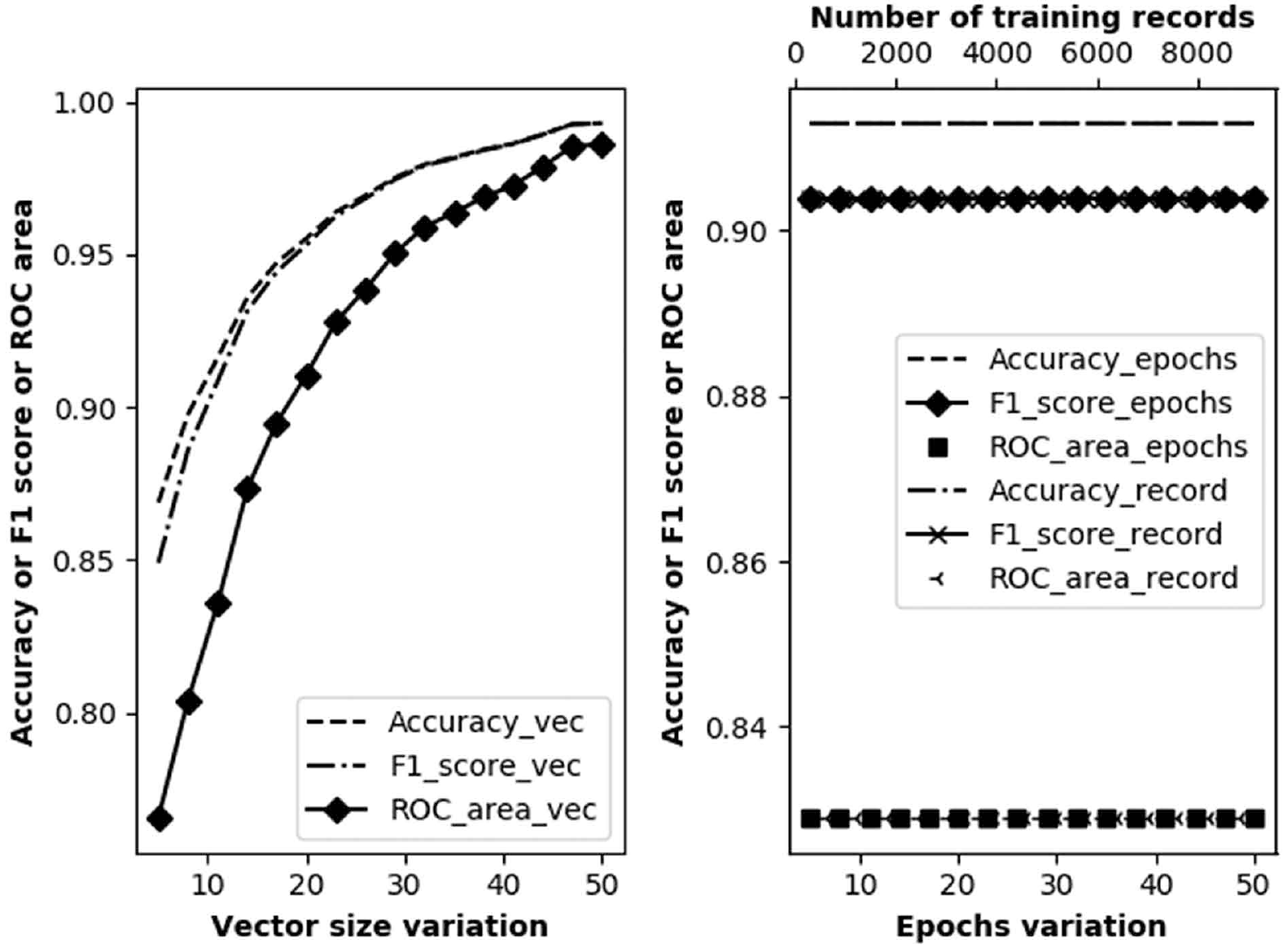}

\caption{Performance of software description similarity detection models with different parameter inputs}
\label{fig:parameter-tuning}
\end{figure*}

\begin{table}
\centering
\caption{Similarity score stats for same software product description pairs}
\label{table:thresholds}
\begin{tabular}{ c|c|c|c|c|c }
\toprule
\textbf{Measure} & $\alpha_{min}$ & $\alpha_{avg}$ & $\alpha_{stddev}$& $\alpha_{max}$&$\hat{\alpha}$ \\ \midrule
\textbf{Value} & 0.9999 & 0.9999 & 9.5305e-08 & 1 & 0.9999 \\ \bottomrule
\end{tabular}
\end{table}

\begin{table}
\centering
\caption{Best performance matrix (Models trained using 10 epochs, vector size 50, training samples 18260)}
\label{tab:evaluation-metrics-values}
\begin{tabular}{ c|c|c|c|c|c }
\toprule
\textbf{Measure} & $Accuracy$ & $Precision$ & $Recall$& $F1\ score$&$ROC\ area$ \\ \midrule
\textbf{Value} & 99.3\%  & 99.24\%  & 98.62\% & 99.3\% & 98.62\% \\ \bottomrule
\end{tabular}
\end{table}

\subsubsection{Results and Observations}
\label{sec:exp-1-results-and-observations}

Fig. \ref{fig:parameter-tuning} shows the performance of software description similarity models (in terms of F1 score and ROC area) with different input parameter combinations. Some of the key observations are as follows:

\begin{enumerate}
    \item The evaluation metrics values remain almost constant on varying the epochs and training dataset size but increase uniformly with the increase in vector size.
    \item The highest evaluation metrics values (viz., accuracy, F1 score, and ROC area) are recorded corresponding to the $\gamma=50$.  Thus, we select the fine-tuned input parameters as $\gamma=50$ and $\beta=10$ for our further experiments.
    \item The highest accuracy, F1 score, and ROC area values achieved by our software description similarity models are 99.3\%, 99.3\%, and 98.62\%, respectively.
\end{enumerate}

Table \ref{tab:evaluation-metrics-values} lists the highest evaluation metrics' values obtained by testing the software description similarity model trained using optimally tuned parameters $\beta=10$ and $\gamma=50$. 

\subsection{Details of the SDEE relational schema}
\label{ap:SDEE-dataset-details}

A brief description of the tables of this schema is as follows:
\begin{enumerate}
\item \emph{release\_info}: This table stores the information about the releases of various OSS repositories. The attributes \emph{repo} and \emph{owner} represent the repository names and the respective repository owner names. The attributes \emph{release\_no}, \emph{release\_date} and \emph{size} represent the release number, deployment date and size of different releases of the repositories respectively.

\item \emph{commit\_stats}: In this table, we store the developer-activity information extracted corresponding to various commits of a repository. The attributes \emph{commit\_id} and \emph{dev\_id}  represents the unique IDentifiers (IDs) assigned to  various commits and developers, respectively. 

For each of the commits stored in the table, we also store the development \emph{effort} and time (\emph{dev\_time}) spent in generating the commit. To compute the commit-wise effort and dev\_time, we employ the use of \emph{basic} and \emph{derived} SDEE metrics, such as SLOC modified by the developer ($SLOC_{modifications}$), developer productivity($P_{d_i}$), skill factor ($P_{d_i}$), etc., discussed in \S\ref{sec:SDEE-metrics}.

\item \emph{release\_effort\_estimate}:
\label{pt:dev-act-fac}
This table stores release-level effort estimates computed by utilizing the SDEE metrics' values present in the commit\_stats table. 
The attributes \emph{min\_release\_ids} and \emph{max\_release\_ids}  represent the IDs of the considered releases corresponding to various \{owner, repo\} pairs of repositories. Similarly, the attributes \emph{start\_release\_date} and \emph{end\_release\_date} represent the considered release period of effort computation, which corresponds to the release dates of the involved releases, and \emph{days} represents the considered time-period interval expressed in the form of days. The attribute \emph{dev\_count} represents the total count of developers contributing to the specific repository in the specified period. 
\item \emph{soft\_desc\_pva\_vec}:
\label{pt:soft-desc-pva-vec}
This table stores the software project descriptions' PVA vectors and the respective cosine similarities with the reference vectors. The \enquote{category} attribute represents the software category as specified on MavenCentral.com.

\end{enumerate}